\documentclass[aps, pra, reprint, amsmath,amssymb,bibnotes, floatfix, showkeys]{revtex4-1}

\usepackage{natbib}
\usepackage{graphicx}
\usepackage{bm}
\usepackage{hyperref}
\usepackage[mathlines]{lineno}

\begin{document}

\preprint{APS/123-QED}

\title{Tracing light propagation to the intrinsic accuracy of space-time geometry}

\author{Mariateresa Crosta}
\affiliation{INAF, Astronomical Observatory of Torino \\ via Osservatorio 20, I-10025 Pino Torinese (TO), Italy} 


\date{\today}

\begin{abstract}
Advancement in astronomical observations and technical instrumentation requires coding light propagation at high level of precision; this could open a new detection window of many subtle relativistic effects suffered by light while it is propagating and entangled in the physical measurements. Light propagation and its subsequent detection should indeed be conceived in a fully relativistic context, in order to interpret the results of the observations in accordance with the geometrical environment affecting light propagation itself, as an {\it unicum} surrounding universe.  One of the most intriguing aspects is the boost towards the development of highly accurate models able to reconstruct the light path consistently with General Relativity and the precepts of measurements. This paper deals with the complexity of such a topic by showing how the geometrical framework of models like RAMOD, initially developed for astrometric observations, constitutes an appropriate physical environment for back tracing a light ray conforming to the intrinsic accuracy of space-time. This article discusses the reasons why RAMOD stands out among the existent approaches applied to the light propagation problem and provides a proof of its capability in recasting recent literature cases.
\end{abstract}

\maketitle


\section{ \label{sec:intro}Introduction}

The treatment of light propagation in time-dependent gravitational fields, as the surrounding universe, is extremely important for astrophysics, encompassing issues from fundamental astronomy to cosmology. Attaining very accurate measurements will allow to observe a new range of subtle physical effects (see, for example, \citet{2001lrevr...2..4..W, 2009IAU...261.1003T, 2006CQGra..23.4853C, 2000tmcs.conf..303K, 1999PhRvD..59h4023K, 2006CQGra..23.4299K, 2001gcit.conf..141B, 2001ApJ...556L...1K, 2007PhRvD..75f2002K,1993PhRvL..70.2217D, 2002PhRvL..89h1601D}). 

Undoubtedly, light tracing should be conceived in a general relativistic framework. Today General Relativity (GR) is the theory in which geometry and physics are joined together in order to explain how gravity works. The trajectory of a photon is traced by solving the relativistic equations of the null geodesic in a curved space-time and, at the same time, the detection process usually takes place in a geometrical environment generated by a n-body distribution as could be that of our Solar System, i.e., by a localized gravitationally-bound distribution of masses moving with their velocities and accelerations. Once the physical set-up is defined, the null geodesic should be made explicit according to the corresponding geometry.

Nevertheless the exact solution of the null geodesic is not generally known. Hence, one has to resort to approximation methods.
Nowadays several conceptual approaches model the light propagation problem in the relativistic context. 
Among them, the post-Newtonian (pN) and the post-Minkowskian (pM) approximations are those mainly used (\citet{1999PhRvD..60l4002K, 2002PhRvD..65f4025K, 2003AJ....125.1580K,2006gr.qc....11078T}; and references therein). 

The peculiar between the two approximations consists in the fact that pN is based on the assumption of an everywhere-weak gravitational field and of slow motion, while the pM only on the weakness of the gravitational field. This can be formalized \citep{1987thyg.book.s6.9D}  by saying that the expansion of the metric tensor can be characterized in powers of a small parameter, 1/c in the pN case \footnote{The pN expansion is valid inside the near-zone, i.e. inside a distance comparable to the wavelength of the gravitational radiation emitted by the bounded system (Fock V., 1959, Pergam Press: Oxford).} and G in the pM approximation (which contains also terms in 1/c). Basically the approximation methods rely upon the idea of being pN as close as possible to the Newtonian theory (absolute time, absolute space, auxiliary Euclidean metric and instantaneous potential) and pM to the Minkowskian interpretation of special relativity (absolute space-time, auxiliary Minkowskian metric and retarded potential) \citep{1987thyg.book.s6.9D}. 

In the framework of the pN approximation of GR, the light path is solved using a matching technique which links the perturbed internal solution inside the near-zone of the n-body system with the external one. Applications of this method are adopted, for example, in \citep{1992AJ....104..897K, 2003AJ....125.1580K, 2003AAp...404..783K}, where the boundary conditions for the null geodesic are fixed by the coordinate position of the observer and imposing the value of $c$ to the modulus of the light ray vector at past null infinity. 
However, the expansion of the metric tensor may not be analytic in higher pN approximations \cite{1987thyg.book.s6.9D}. The second pN approximation in propagation of light  was investigated by Brumberg in \cite{1991ercm.book.....B}. Advancement on second order pN equations of light propagation are made in \citet{2005gr.qc....10074X} or, more recentely, in  \citet{2010IAUS..261..103T}. Moreover, the analytical solution for light propagation in the gravitational field of one spherically symmetric body in the framework of post-post-Newtonian (ppN) development has been re-analyzed \cite{2010CQGra..27g5015K}. Analytical formulas have been compared with high-accuracy numerical integrations of the geodetic equations, the latter based on an approximation of the exact analytical solution. The authors found a deviation between the standard post-Newtonian approach and the high-accuracy numerical solution of the geodetic equations  which amounts to 16 $\mu$-arcsecond. As reported by \citet{2010CQGra..27g5015K},  detailed analysis has shown that the error is of post-post-Newtonian order in contradiction with the analytical estimate. To clarify the problem, standard post-Newtonian and post-post-Newtonian solutions have been derived by retaining terms of relevant analytical orders of magnitude. For each individual term in the relevant formulas exact analytical upper estimates have been found and a compact analytical solution for light propagation is derived up to 1 $\mu$-arcsecond accuracy. The ÔenhancedÕ post-post-Newtonian terms calculated by Klioner and Zschocke, are confirmed recently by Teyssandier \citep{2010gr.qc....1012.5402v1T}

On the other hand, \citet{1999PhRvD..60l4002K}, using the pM approximation, solve Einstein's equations in the linear regime by expressing the perturbed part of the metric tensor in terms of retarded Lienard-Weichert potentials. Later, \citet{2002PhRvD..65f4025K} included all the relativistic effects related to the gravitomagnetic field produced by the translational velocity/spin-dependent metric terms. In these works a special technique of integration of the equation of light propagation \citep{1997JMP....38.2587K} with retarded time argument is extended both outside and inside of a gravitating system of massive point-like particles moving along arbitrary world lines.
The null geodesic is rewritten as function of two independent parameters, one related to the Euclidean scalar product between the light ray direction and its unperturbed trajectory, and the other derived as the constant impact parameter \footnote{With respect to the origin of the coordinate.} of the unperturbed trajectory of the light ray. Then, the light trajectory is again traced as a straight line plus integrals, each containing the perturbations encountered, from a source at an arbitrary distance from an observer located within the n-body system. 

As a consequence, one more difference between the pN/pM approximations appears in the computation of the light deflection contributions: in the pN scheme this is done by combining the external and internal solution with the technique of asymptotic matching, while the pM method utilizes in addition a semi-analytical integration of the equation of light propagation from the observer to the source with respect to the retarded time considered as argument in the metric tensor. The relativistic perturbations remain expressed in terms of integrals (i.e. a semi-analytical solution) which can be solved analytically in specific cases (i.e. not expressed as a general solution of the differential equation, but as particular solutions for some effects on the light propagation) once the motion of the gravitating bodies is prescribed.  The pM approximation is also used by \citet{2008CQGra..25n5020T} that found a way to bypass the solution of the null geodesic  in form of differential equations. In this work the authors elaborate the light trajectory with the powerful tool of the Time Transfer Function \cite{2004CQGra..21.4463L}, worked out throughout a recursive procedure for expanding the Synge's World Function, a covariant  method that is conceptually much closer to that of RAMOD, the approach discussed in the present work.

RAMOD stands originally for Relativistic Astrometric MODel, conceived to solve the inverse ray-tracing problem in a general relativistic framework not constrained by a priori approximations. RAMOD is, actually, a family of models of increasing intrinsic accuracy all based on the geometry of curved manifolds (\citep{2004ApJ...607..580D,2006ApJ...653.1552D} and reference there-in) where, as this paper proves, light propagation can be expressed in a general relativistic context. 
An analytical solution of the null geodesic exists in the case of the Parametrized-Post-Newtonian Schwarzschild metric \citep{1973grav.book.....M,2001lrevr...2..4..W}, which, just considering the spherical mass of the Sun, can be implemented for the Gaia mission (Esa, \citet{2005tdug.conf.....T}) and could prove the dilaton-runaway scenario \citep{2003AAp...399..337V}.
At present, the RAMOD full solution requires the numerical integration of a set of coupled non linear differential equations, called ``master equations'',  for tracing back the light trajectory to the initial position of the star, and which naturally entangles the contributions of the curvature of the background geometry.
A solution of this system of differential equations contains all the relativistic perturbations suffered by the photon along its trajectory
due to the intervening gravitational fields. The boundary condition are usually fixed by the required physical measurement \citep{2003CQGra..20.4695B}. In this context, a semi-analytical solution was found by \citet{2006CQGra..23.5467D}.

This work shows how RAMOD can naturally recover the seminal work of \citet{1999PhRvD..60l4002K} and \citet{2003AJ....125.1580K}.  While the third one stems from the general formulation of the second by conforming to its basic conception \citep{1992AJ....104..897K}, RAMOD responds to an inherently different strategy. As far as the invers light ray  tracing  problem is concerned, in the Kopeikin and Sh\"afer and Klioner approaches the light ray is reconstructed as a sum of terms which allows for a direct evaluation of the individual relativistic effects, induced by the gravitating bodies, suffered by the light on its way to the observer.  
RAMOD, instead, aims to determine a full solution for the light trajectory which naturally include, in a curved space-time, all the individual effects; the latters are somewhat hidden in the covariant formalism of this approach and directly contribute to the solutions of the master equations.  Evidently, any specific effect one is interested to explore can be independently deduced from our formalism as a branch output, once we adapt the model to a required specific case. 
A proof of this is given in \citet{2010A&A...509A..37C}, where a first comparison between RAMOD and GREM (Gaia RElativistic Model, \citep{2003AJ....125.1580K}, which employes the pN formulation) was carried out via the extrapolation of the aberrational term in the "local" light direction, i.e., at the observer. 

Now, if we consider a gravitational field generated by a n-body system, the required accuracy in back tracing the light ray should be consistent with the geometry induced by the system itself.  The accuracy is gauged by the virial theorem, i.e., by the ratio $v/c$, where $v$ is the typical velocity of a body of the system. In this respect, RAMOD3 \citep{2004ApJ...607..580D} is developed up to the order $(v/c)^2$, namely, it is based on a static geometry of the space-time, whereas the most recent RAMOD4  \citep{2006ApJ...653.1552D} is the extension of the predecessor to include a dynamical geometry (i.e., all metric terms up to the order $(v/c)^3$), which is accomplished by including also the vorticity contribution proportional to the off-diagonal term $g_{0i}$ of the metric. Both models introduce the retarded contributions to the metric due to the gravitational influence of the sources on the light path, according to the natural space-time structure. 
In section \ref{sec:sec1}, we discuss how to build an appropriate geometrical set-up in order to handle light propagation
through such a space-time. In section \ref{sec:sec2} we present the building process of the differential equations, i.e., the master equations, from which  the light trajectory from an observer, located within the n-body system, to a distant star can be reconstructed. We distinguish two cases:  (i) the static space-time (RAMOD3), and  (ii) its dynamical extension (RAMOD4). An application of RAMOD3 to the euclidean metric is presented in section \ref{sec:sec3}, while section \ref{sec:sec4} shows how this last computation allows to parametrize the RAMOD3 master equation as done in \citep{1999PhRvD..60l4002K}. Finally, section \ref{sec:sec5} shows how the velocity of the bodies can enter the metric term simply as a "retarted distance" corrections in RAMOD3-like models. 

\section{\label{sec:not}Notations}
In this paper the following notations will be used:
\begin{itemize}
\item components of vectorial quantities are indicated with indexes (no bold symbols), where the latin index stands for 1,2,3 and the greek ones for 0,1,2,3; 
\item regular bold indicates four-vector (e.g. ${\mathbf u}$);
\item indexes are raised and lowered with the metric $g_{\alpha \beta}$; 
\item $n^i n_i$ or $n^{i} n^{i}$ stands for the scalar product with respect to the euclidean metric $\delta_{ij}$, whereas $l^{\alpha} l_{\alpha}$ with respect the metric $g_{\alpha \beta}$;  
\item $c$ indicates the fundamental speed (equal to the speed of light in vacuum);
\item A quantity like $ \dot{\Upsilon}^\alpha$, represents the tangent vector to a curve $\Upsilon$;
\item $P(\mathbf{u})_{\alpha \beta}$ represents the operator which projects orthogonally to its argument. 
\item $x'^{\alpha} $ or $x^{\alpha} $ indicate generic four-coordinates, whereas $\xi^{\alpha} $ the Lie-transported ones;
\item dot applied over coordinates means derivative with respect to the coordinate time; 
\item partial derivative with respect to the coordinates are usually indicated by a comma. Instead, the symbol $\bf{\partial}$  is adopted in cases where the formulas need to be more explicit and avoid confusion; covariant derivative are indicated with ${\mathbf \nabla}$.
 \end{itemize}

\section{\label{sec:sec1} Setting up the geometry}

The RAMOD framework is based on the small curvature limit, for which the curvature of the background geometry is sufficiently small to neglect non linear terms. The degree of this approximation has in turn to be specialized to the particular case one wishes to model.
The first step is to identify the gravitational sources and then fix the background geometry.  

Let the space-time be generated by a weakly relativistic gravitationally bound system; in this case the metric is
\begin{equation}
\label{eq:met}
g_{\alpha \beta }= \eta_{\alpha \beta} + h_{\alpha \beta} + O(h^2),
\end{equation}
where  ${\eta_{\alpha\beta}}$ is the flat Minkowskian metric and the ${h_{\alpha\beta}}'s$ describe effects generated by the bodies of the system and are {\it small} in the sense that $|h_{\alpha\beta}|\ll 1$. From the virial theorem all forms of energy density within the system must not exceed the maximum amount of the gravitational potential in it, say, $U$. So, the energy balance requires that $|h_{\alpha \beta}|\le U/c^{2} \sim v^{2}/{c}^{2}$, where $v$ is the average relative velocity within the system~\footnote{For a typical velocity  $\sim 30$ km/s, $(v/c)^2 \sim$ 
1 milli-arcsec}. Since the latter is weakly relativistic~\footnote{This means that the length scale of the curvature is everywhere small compared to the typical size of the system, which implies also that $v^{2}/c^{2} << 1$}, the ${h_{\alpha\beta}}'s$ are at least of the order of $(v/c)^2$ and the level of accuracy, to which is expected to extend the calculations, is fixed by the order of the small quantity $\epsilon\sim (v/c)$. In summary, the perturbation tensor $h_{\alpha \beta}$ contributes with terms even in $\epsilon$ to $g_{ 00 }$ and $g_{ ij}$ (lowest order $\epsilon^{2}$) and with terms odd in $\epsilon$ to $g_{0i}$ (lowest order $\epsilon^{3}$, \citealt{1973grav.book.....M}); its spatial variations is on the order of $|h_{\alpha\beta}|$, while its time variation is on the order of $\epsilon| h_{\alpha\beta}|$. Clearly the metric form (\ref{eq:met}) is preserved under to infinitesimal coordinate transformations of order $|h|$.

Under these hypothesis, in order to describe the space-time evolution of the system, let us introduce a family of physical observers, i.e. a time-like congruence of curves, and consider its vorticity, which measures how a world-line of an observer rotates around a neighboring one. If an open set of the space-time manifold admits a vorticity-free congruence of lines, then it can be {\it foliated} (Frobenius theorem, \citep{1990recm.book.....D, 2010ToM.book.....D}). 

\begin{figure}[htp] 
\begin{centering}
\includegraphics[width= 0.8\columnwidth]{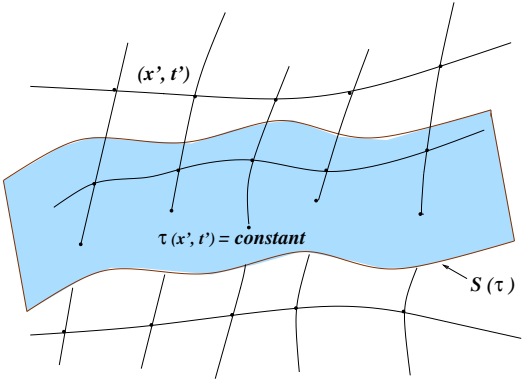}
\caption{A coordinate grid $({\bf x'},t')$ and the hypersurface $\tau ({\bf x'},t') =$ constant.} 
\label{fig:griglia} 
\end{centering}
\end{figure}

Given a general coordinate space-time representation $(x^{'i},t')_{i=1,2,3}$, the associated metric being $g'_{\alpha\beta}(x',t')$, there exists a family of three-dimensional hypersurfaces  $S(\tau)$, or {\em slice}, described by the equation:
\begin{equation}
 \tau(x',t')= {\rm constant}, \label{slice} 
\end{equation} 
where $\tau(x',t')$ is a real, smooth and differentiable function of the coordinates.  A space-like foliation implies that a unitary one-form $u'_{\alpha}$ exists which is everywhere proportional to the gradient of $\tau$ (figure \ref{fig:griglia}), namely \citep{1990recm.book.....D,1973grav.book.....M}:
\begin{equation} u'_{\alpha}=- ( \tau_{, \alpha '}) {\rm e}^{\psi '} 
\label{congru1} 
 \end{equation} where
\begin{equation} 
g'^{\alpha \beta} u'_{\alpha} u'_{\beta}= - 1,
\end{equation} 
and
\begin{equation} 
{\rm e}^{\psi '}=[-g'^{ \alpha \beta} (\tau_{, \alpha'})(\tau_{,  \beta '})]^{-1/2} \label{factor1}.
\end{equation}

Evidently the curves tangent to the vector field, i.e. $u^{'\alpha}=g'^{\alpha\beta}u'_\beta$, form a congruence of time-like lines $C_{\mathbf u'}$, everywhere orthogonal to the slices $S(\tau)$. 
 Let us compute the vorticity as the tensor quantity \citep{1990recm.book.....D} 
\begin{equation} 
\omega_{\alpha \beta }\equiv {P'}_\alpha{}^\rho{P'}_\beta{}^\sigma u'_{[\rho,\sigma]},  \label{eq:vort}
\end{equation}
where square brackets mean anti-symmetrization and
\begin{equation} 
{P'}_{\alpha \beta } (\mathbf {u}')= {g'}_{\alpha \beta} + {u'}_{\alpha } {u'}_{\beta } \label{proj1}
\end{equation}
is a projector operator on  $S(\tau)$. Since $ u'_{[\alpha , \beta ]}$ results in $u'_{[\alpha } \partial_{\beta ]}\psi '$, we find that the congruence $C_{\mathbf u'}$ is vorticity-free only up to $(v/c)^2$ order; namely,
\begin{equation} 
\omega_{\alpha \beta }={P'}_\alpha{}^\rho{P'}_\beta{}^\sigma u'_{[\rho }
\partial'_{\sigma ]}\psi ' = O \left(h_{0i} \right).  \label{eq:vort}
\end{equation} 
Then, we may affirm that considering only terms up to $\epsilon^{2} \sim (v/c)^2$, it is the same as assuming that structure of space-time is static. 
So, does the one-form ${u'}_{\alpha}$ have a physical meaning? Let us introduce a new coordinate system $(x^0,x^i)$ such that:  
\begin{equation}
\label{eq:coor1}
\begin{cases}
 x^0 =  \tau ({\vec x'}, t') \cr x^i = x^i ({\vec x'}, t') & {}
\end{cases}
\end{equation}
where $x^i$ are the space-like coordinates defined on each slice.
From (\ref{eq:coor1}), the congruence of normals now have the following tangent vector field:  
\begin{equation}
\begin{cases}
u_{\alpha}(x^0,x^i)= -\delta^{0}_{\alpha} {\rm e}^{\psi} \cr 
u^{\alpha}(x^0,x^i)= -g^{0 \alpha}{\rm e}^{\psi} = 
{\displaystyle \frac {d x^\alpha}{d \sigma}} & {}
\end{cases}
\label{eq:u}
\end{equation}
where $\psi=\psi(x^{0}, \vec x)$ is a unitary function  ($u^\alpha u_\alpha=-1$), $\sigma$ is the parameter on the normals, and
\begin{eqnarray}
&&u_{\alpha}(x^0,x^{i})=\frac{\partial x^{' \beta}}
{\partial x^{\alpha}} u^{'}_{\beta}({\bf x'}, t') \nonumber\\
 &&= -\frac{\partial \tau}{\partial x^{\alpha}}{\rm e}^{\psi'({\bf
x'}, t')}=-\delta^{0}_{\alpha}{\rm e}^{\psi(x^0,
x^{i})}. \end{eqnarray}
Clearly it is 
\begin{equation} 
{\rm e}^{\psi}=\frac{d\sigma}{d\tau}= \Big(-g^{00}\Big)^{-1/2}\label{eq:epsi}
\end{equation} 
which means that the parameter $\sigma$ runs uniformly with the coordinate time $\tau$.
The congruence $C_{\mathbf { u}}$, however, does not Lie-transport the spatial
coordinates $x^i$.
From (\ref{eq:coor1}) and (\ref{eq:ucheck}), the line element can be cast into the form:
\begin{equation} 
ds^2=-(Ndx^0)^2+g_{ij}(dx^i-N^idx^0)(dx^j-N^jdx^0)
\label{eq:lineel}
\end{equation} 
where $N\equiv \pm 1/ u^0$ and $N^i={{u}^i}/{{u}^0}$
are termed respectively {\it lapse} and {\it shift} functions (or factors) \citep{1973grav.book.....M}.  
In fact, choosing at time $\tau_1$, say, an arbitrary event $\cal{P}$, labeled by the coordinates $(\tau_1, x^i_1)$,
the unique normal through that point will intersect the slice
$S(\tau_1+\Delta\tau)$ at a point $\cal{P'}$ with spatial coordinates which
are shifted from the initial ones by the amount (see appendix A):
\begin{equation} 
\label{eq:shift}
\Delta x^i=\int^{\Delta\tau}_0 N^i(\tau')d\tau'. 
\end{equation}  

The spatial coordinate transformation to be applied, in order to  make the shift factor vanish and assure that the congruence of
normals will now Lie-transport the spatial coordinates, 
is:
\begin{equation}
\label{eq:newcoor} 
\begin{cases}
d\xi^i=dx^i-N^idx^0 \cr
d\xi^0= dx^0=d\tau, 
\end{cases}
\end{equation} 
in which we impose:
\begin{equation}
\tilde{u}^{\alpha}({\xi}^0,{\xi}^i)=
\frac{\partial {\xi}^{\alpha}}
{\partial x^{\beta}}{u}^{\beta} \equiv {\rm e}^{\phi} 
\delta^{\alpha}_{0}.
\label{eq:impo}
\end{equation} 
Under the above transformation, we clearly have for $\alpha=0 $:
\[
\left(\frac{\partial {\xi}^{0}}{\partial \tau } {g}^{00}+
\frac{\partial {\xi}^{0}}
{\partial x^i}{g}^{0i}\right){\rm e}^{\psi}=-{\rm e}^{\phi}
\]

\[
\Longrightarrow {\rm e}^{\phi}= \sqrt{-{g}^{00}}
\]
from (\ref{eq:epsi}); and for  $\alpha=i $ 
\[
\frac{\partial {\xi}^i}{\partial \tau } u^{0}+
\frac{\partial {\xi}^i}
{\partial x^j}{u}^{j}=
\frac{\partial {\xi}^i}{\partial \tau}{g}^{00}+
\frac{\partial {\xi}^i}{\partial x^{j}}{g}^{0j}= 0.
\]

But it is:
\[
\tilde{g}^{0 \alpha}= \frac{\partial {\xi^0}}{\partial x^{\rho}} 
\frac{\partial {\xi^{\alpha}}}{\partial x^{\sigma}}
{g}^{\rho \sigma}
=\frac{\partial {\xi}^{\alpha}}{\partial x^{\sigma}}
{g}^{0 \sigma}
\]
which implies
\begin{equation}
\tilde{g}^{00}={g}^{00} 
\label{eq:congoo}
\end{equation}
\begin{equation} 
\tilde{g}_{0i}= \tilde{g}^{0i}= \frac{\partial {\xi}^i}{\partial \tau} 
{g}^{00}+ \frac{\partial {\xi}^i}{\partial x^j} 
{g}^{0j}= 0.
\label{eq:cong0i}
\end{equation}
 
Then the vector field tangent to the congruence $C_{\tilde {\mathbf u}}$ which transports the spatial coordinate is: 
\begin{equation}
\label{eq:utilde}
\begin{cases}
\tilde u^\alpha(\xi^0, \xi^i) =  {\displaystyle \frac{d\xi^\alpha}{d\sigma}}= {\rm e}^{\phi(\tau, \vec \xi)}\delta^\alpha_0 \cr  \tilde u_\alpha(\xi^0,\xi^i)={\rm e}^\phi g_{0\alpha}= -{\rm e}^\psi {\displaystyle \frac{\partial\tau}{\partial\xi^\alpha}} & {}.
\end{cases}
\end{equation}

Let us summarize. First, the vorticity is proportional to the $g_{0 i}$ term of the metric, the lowest order established by equation~\ref{eq:vort};
second,  we must bear in mind that, even if metric form (\ref{eq:met}) holds up to infinitesimal coordinate transformations, {\em the condition
that the shift factor is zero is preserved under the above transformation only to the order $\epsilon^2$}. In fact, to the order of $\epsilon^3$, the terms $g'_{0i}$ will not remain zero if they were so initially. 
We may affirm that considering only terms up to $\epsilon^{2} \sim (v/c)^2$, it is the same as assuming that structure of space-time is static, since a  static space-time is a stationary spacetime in which the timelike Killing vector field has vanishing vorticity, or equivalently (by the Frobenius theorem) is hypersurface orthogonal, i.e. admits a family of spacelike surfaces of constant time (\citep{2004ApJ...607..580D,1990recm.book.....D,1973grav.book.....M}). The parameter $\sigma$ along the Killing congruence $C_{\tilde {\mathbf u}}$, such that $\tilde{u}^\alpha=d{\xi}^\alpha/d\sigma$, is the proper time of the physical observers who Lie-transport the spatial coordinates.  Any hypersurfaces, at each different coordinate time $\tau $, can be considered the rest space of the observer $\tilde {\mathbf u}$. One can require that the world line of the center of mass of the body's  system, chosen as origin of coordinates of the gravitational bounded system, belongs to this congruence while the world lines of the bodies would differ from the curves of the congruence by an amount which depends on the local spatial velocity relative to the center of mass; however, at the $\epsilon^2$ order, i.e. neglecting the relative motion of the gravity sources, one can assume that also their world lines belong to that congruence and the geometry that each photon feels is, then, identified with the metric $\tilde{g}_{\alpha\beta}$.  
To the order of $\epsilon^3$ the shift factor should be retained and so also terms proportional to $h_{0i}$ (eq.~\ref{eq:vort}). Since we require that the background geometry is weakly relativistic, we can assume that the vorticity is free up to the $\epsilon^3 $ order only locally; then the space-time still admits a family of hypersurfaces $\tau=$constant with normals forming a congruence of curves with tangent field $\mathbf {u}$ given by  (\ref{eq:u}). The given coordinate system $(x^i,\,\tau)$, where $x^i$ are fixed on each slice, is still centered to the baricenter of the n-body system, but, on the other hand, each slice is not the local rest-space of the observer $\mathbf u$. These normals do not in general Lie-transport the 
spatial coordinates $x^i$; instead they vary along the normals according to the shift law (\ref{eq:shift}). 
Any of these observers can be considered at rest with respect to the coordinates $x^i$ {\em only locally}, and for this reason $\mathbf u$ is called  {\em local barycentric observer}. 

\section{\label{sec:sec2}From the null geodesic to the master equations}
\begin{figure}[htp] 
\begin{centering}
\includegraphics[width=0.8\columnwidth]{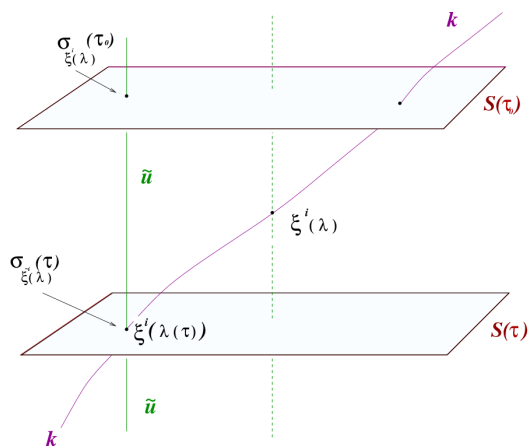}   
\caption{The point $\xi^i(\lambda(\tau))$ belongs to the null
geodesic but also to the unique normal to the slice $S(\tau)$ crossing it with a value of the parameter
$\sigma_{\xi^i(\lambda)}(\tau)$.} \label{fig:fig/figtesi4}
\end{centering}
\end{figure}

\begin{figure}[htp] 
\begin{centering}
\includegraphics[width= 0.8\columnwidth]{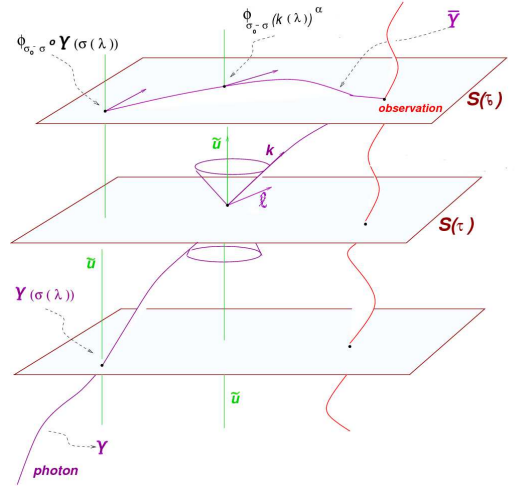}   
\caption{The curve $\bar\Upsilon$ is the image of the null geodesic 
$\Upsilon_k$ under the diffeomorfism $\phi_{\Delta \sigma}$. } 
\label{fig:fig5} 
\end{centering}
\end{figure}

In the following sections we drop the tilde for the metric.
A photon travelling from a distant star to an observer, located
within the bounded system, would see the space-time as a time
development of the $\tau=$~constant slices. 
Let us now consider a null geodesic $\Upsilon_k$ with tangent
vector field $k^\alpha\equiv d\xi^\alpha/d\lambda$ which
satisfies the following equations:
\begin{eqnarray} 
\label{eq:geodes} 
k^\alpha k_\alpha&=&0\\ 
\frac{dk^\alpha}{d\lambda}+{\Gamma}^{\alpha}_{\rho\sigma} 
k^\rho k^\sigma&=&0;\label{eq:null-geodesic} 
\end{eqnarray} 
here $\lambda$ is a real parameter on $\Upsilon_k$ and
$\Gamma^{\alpha}_{\rho\sigma}$ are the connection coefficients
of the given metric.  Assume that the trajectory starts at a point
$\cal{P}_\ast$ on a slice $S(\tau_*)$ (say) and with spatial coordinates
$\xi_*^i$.  
The null geodesic crosses each slice $S(\tau)$ at
a point with coordinates $\xi^i=\xi^i(\lambda(\tau))$; but
this point also belongs to the unique normal to the slice $S(\tau)$,
crossing it with a value of the parameter
$\sigma=\sigma(\xi^i(\lambda),\,\tau) \equiv \sigma_{\xi^i(\lambda)}(\tau)$ (figure \ref{fig:fig/figtesi4}). 
Let us now define the one-parameter local diffeomorphism \citep{1990recm.book.....D}: 
\begin{equation} 
\phi_{\Delta\sigma}\equiv\phi_{(\sigma_{\xi^i(\lambda)}
(\tau_0)- \sigma_{\xi^i(\lambda)}(\tau))}\,:\,\Upsilon_k\cap S(\tau) \to S(\tau_0)
\label{eq:diff} 
\end{equation} 
which maps each point on the null geodesic $\Upsilon_k$ with the point 
on the slice $S(\tau_o)$ which one gets to by moving along the unique normal 
through the point $\Upsilon_k(\lambda)\cap S(\tau)$, by a parameter 
distance $\Delta\sigma=\sigma_{\xi^i(\lambda)}(\tau_o)-
\sigma_{\xi^i (\lambda)}(\tau)$ (figure \ref{fig:fig5}). 
Since the spatial coordinates $ \xi^i$ are Lie-transported along the 
normals to the slices, then the points in $S(\tau_0)$, which are images of 
those on the null geodesic under $\phi_{\Delta\sigma}$,
have coordinates $(\phi_{\Delta\sigma}(\Upsilon_k((\lambda)
\cap S(\tau)))^i=\xi^i$.
The  curve in $S(\tau_0)$ which is the image of $\Upsilon_k$ 
under $\phi_{\Delta\sigma}$, is: 
\begin{equation} 
\bar{\Upsilon}\equiv\phi_{\left(\sigma_{\xi^i(\lambda)}(\tau_o)- 
\sigma_{\xi^i(\lambda)}(\tau)\right)}\circ\Upsilon_k ,
\label{eq:curveim} 
\end{equation} 
with tangent vector \citep{1990recm.book.....D}
\begin{equation} 
\dot{\bar{\Upsilon}}^\alpha=\dot{\Big(\phi_{\Delta\sigma}^*\circ k\Big)}^\alpha= \frac{\partial\xi^\alpha(\sigma(\tau_o))}{\partial\xi^\beta 
(\sigma(\tau))}k^\beta\equiv\ell^\alpha. \label{eq:tangveim} 
\end{equation} 

This coincides with the projection operation on the rest-space of the observer $\tilde{u}$ in any point of the mapped trajectory on the silde:
\begin{equation} 
\ell^\alpha=P(\tilde u)^{\alpha}_{\beta} k^\beta\,, 
\label{eq:ell} 
\end{equation} 
hence the curve $\bar{\Upsilon}$ is the spatial projection of the null geodesic on the slice $S(\tau_o)$ at the time of observation;
this curve will be denoted as $\bar{\Upsilon}_\ell$ and is naturally parameterized by $\lambda$.  Then from equation (\ref{eq:tangveim}) by using (\ref{eq:utilde}) (where $\tilde u^\alpha \tilde u_\alpha=-1$) or (\ref{eq:ell}), it follows that:
\begin{equation} 
\ell^\alpha=k^\alpha+\tilde u^\alpha (\tilde u_\beta k^\beta)\,. 
\label{eq:ellsplit} 
\end{equation} 
Clearly by setting $\alpha=0$ and from equation (\ref{eq:utilde}) we deduce $\ell^0=0$; moreover, $\ell^\alpha$ is space-like as expected as it is a projection on slice $S(\tau_0)$ and lies everywhere in it.  
Since each point of $\bar{\Upsilon}_\ell$ is the image under the diffeomorphism $\phi_{\Delta\sigma}$ of a point $\Upsilon_k\cap
S(\tau)$, it is more convenient to label the points of $\bar{\Upsilon}_\ell$ with the value of the parameter
$\sigma_{\xi^i(\lambda)}(\tau)$ which, as we have already said, uniquely identifies that point on the normal to the slice $S(\tau)$
which contains it.  Hence, being
\begin{equation}
 d\sigma=-(\tilde u_\alpha k^\alpha) d\lambda\,, 
\label{eq:siglam} 
\end{equation} 
we define the new tangent vector field:  
\begin{equation} 
\bar\ell^\alpha\equiv \frac{d\xi^\alpha}{d\sigma_{\xi^i(\lambda)}}
= -\frac{\ell^\alpha}{\Big(\tilde u^\beta k_\beta\Big)}.
\label{eq:ellbar} 
 \end{equation} 
In the same way we denote
\begin{equation}
\label{eq:kappabar} 
\bar k^\alpha\equiv -\frac{ k^\alpha}{(\tilde u_\beta k^\beta)} 
\end{equation} 
so that  
\begin{equation} \bar k^\alpha=\bar\ell^\alpha+\tilde u^\alpha 
\label{eq:kapsplit} 
\end{equation} 
which implies: 
\begin{equation} \bar\ell^{\alpha } \bar\ell_{\alpha}=1. 
\label{eq:ellun}  
\end{equation}  

Note that the parameter $\sigma_{{\xi}^i(\lambda)}$ is also the one which makes the tangent to the curve $\bar{\Upsilon}$ unitary. To short the notations, in what follows we denote $\sigma_{{\xi}^i(\lambda)}$ as $\sigma$.  We can now write the differential equation which is satisfied by the vector field {\boldmath$\bar\ell$}. From (\ref{eq:siglam}) and (\ref{eq:kappabar}), equations (\ref{eq:null-geodesic}) becomes (see appendix \ref{app:appB}):
\begin{eqnarray} 
\frac{d \bar{\ell}^{\alpha}}{d\sigma}&+& \frac{d \tilde u^{\alpha}}{d\sigma}- 
(\bar{\ell}^{\alpha}+\tilde u^{\alpha})(\bar{\ell}^{\beta} {\tilde u}^{\tau} \nabla_{\tau} {\tilde u}_{\beta}+ 
\bar{\ell}^{\beta}\bar{\ell}^{\tau}\nabla_{\tau}\tilde u_{\beta})\nonumber\\ 
&+& \Gamma^{\alpha}_{\beta \gamma}(\bar{\ell}^{\beta}+\tilde u^{\beta})
(\bar{\ell}^{\gamma}+\tilde u^{\gamma})=0\, . 
\label{eq:geoC} \end{eqnarray}  
In this equation the quantity
$\bar{\ell}^{\beta}\bar{\ell}^{\tau}\nabla_{\tau}\tilde u_{\beta}$ can be written explicitly in terms of the expansion 
$\Theta_{\rho \sigma}$ of $C_{\mathbf {\tilde u}}$ \cite{2004ApJ...607..580D}, as:
\begin{eqnarray} 
\frac{d \bar{\ell}^{\alpha}}{d\sigma}&+& \frac{d \tilde u^{\alpha}}{d\sigma}- 
(\bar{\ell}^{\alpha}+\tilde u^{\alpha})(\bar{\ell}^{\beta}\dot{\tilde u}_{\beta}+ 
\Theta_{\rho \sigma} \bar{\ell}^{\rho}\bar{\ell}^{\sigma})\nonumber\\ 
&+&  \Gamma^{\alpha}_{\beta \gamma} (\bar{\ell}^{\beta}+\tilde u^{\beta})
(\bar{\ell}^{\gamma}+\tilde u^{\gamma})=0 
\label{eq:geoexp} \end{eqnarray} 
where $\Theta_{\rho\sigma}=P^\alpha_\rho P^\beta_\sigma 
\nabla_{(\alpha}\tilde u_{\beta)}$ \cite{1990recm.book.....D}. Since the only non vanishing components
of the expansion are $\Theta_{ij}=(1/2)\partial_0 h_{ij}$, the
expansion vanishes identically as consequence of the assumption to
neglect time variations of the metric.  From this condition equation, after some algebra, (\ref{eq:geoexp}) becomes to the required
order (see appendix \ref{app:appC}):
\begin{equation} 
\frac{d \bar{\ell}^{\alpha}}{d\sigma}+ \frac{d \tilde u^{\alpha}}{d\sigma}- 
(\bar{\ell}^{\alpha} + \tilde u^{\alpha})(\bar{\ell}^{\beta}\dot{\tilde u}_{\beta}) + 
 \Gamma^{\alpha}_{\beta \gamma} (\bar{\ell}^{\beta}+\tilde u^{\beta})
(\bar{\ell}^{\gamma}+\tilde u^{\gamma})=0. 
\label{eq:geowexp} 
\end{equation}
If $\alpha=0$, equation (\ref{eq:geoC}) leads to
$\dfrac{d\bar\ell^0}{d\sigma}=0$ assuring that condition $\bar\ell^0=0$
holds true all along the curve $\bar{\Upsilon}$; if $\alpha=k$
equation (\ref{eq:geoC}) gives the set of differential equations 
that we need to integrate to identify the star or the physical effects related to light propagation (see details in appendix \ref{app:appD},
\cite{2003PhDTh...MRA...C} and \cite{2004ApJ...607..580D}):
%

\begin{eqnarray}
\frac{d\bar{\ell}^{k}}{d\sigma}&=& 
- \bar{\ell}^{k} \left(\frac{1}{2}
\bar{\ell}^{i}h_{00,i}\right) - \delta^{k s} \left( h_{s j, i} 
-\frac{1}{2} h_{ij, s}\right)\bar{\ell}^{i}\bar{\ell}^{j} \nonumber \\
&&+ \frac{1}{2}\delta^{ks} h_{00, s}. \label{eq:geodint}
\end{eqnarray}
 
These equations determine light propagation in the static case, here and after called RAMOD3 master equations. Let us remark, that this definition does not stand for the ``master equation`` used in classical or quantum physics; in this context it represents a set of first-order  nonlinear differential equation describing evolution of the spatial light direction components according to the prescribed geometry.
In the dynamical case, any ``local`` observers $\mathbf u$ which are at rest with respect to the coordinates $x^i$ locally, once intersected by the null ray will {\it see} the light signal along a spatial direction {\boldmath $ \ell$} in his rest space given by  (more details can be found in \citet{2006ApJ...653.1552D})
\begin{equation}
\ell^\alpha=P^{\alpha}_{\beta} (u) k^\beta(\tau)
\label{eq:hatell}
\end{equation}
where $P(u)^\alpha{}_\beta=\delta^\alpha_\beta+ u^\alpha u_\beta$ this time is the operator which projects orthogonally to the  $\mathbf u$'s.  Following the same reasoning of the static case, we parameterize the curve $\Upsilon$ with the parameter $\sigma$ which makes unitary the locally projected vector field {\boldmath$ \ell$} which we term again {\boldmath $\bar\ell$}, so $\bar\ell_\alpha\bar\ell^\alpha=1$. 
The differential equation of the null ray, written in terms of $\bar\ell^{\alpha}$, is still given by (\ref{eq:geoC}), the only difference being that $ h_{0i}\not=0$, $h_{00,0}\not= 0$, and $h_{ij,0}\not= 0$.  After some algebra, equation (\ref{eq:geoC}) now reads:
 
\begin{eqnarray}
\frac{d\bar\ell^0}{d\sigma}&-&\bar\ell^i\bar\ell^j h_{0j,i}-\frac{1}{2}  h_{00,0}=0 \label{eq:diffeq0}\\
&{}&\nonumber\\
\frac{d\bar\ell^k}{d\sigma}&-&\frac{1}{2} \bar\ell^k\bar\ell^i\bar\ell^j h_{ij,0}+\bar\ell^i\bar\ell^j\left( h_{kj,i}-\frac{1}{2} h_{ij,k}\right)\label{eq:diffeqk}\\
&+&\frac{1}{2}\bar\ell^k\bar\ell^i  h_{00,i}+ \bar\ell^i\left(h_{k0,i}+h_{ki,0} - h_{0i,k}\right) \nonumber \\
&-&\frac{1}{2} h_{00,k} -\bar \ell^{k} \bar \ell^{i} h_{0i, 0} + h_{k0,0} =0. \nonumber 
\end{eqnarray}
These equations are named RAMOD4 master equations. Note that in RAMOD4, with respect to RAMOD3, we have one more differential equation for the $\bar\ell^{0}$ component. Term like $h_{k0,0}$ is redundant at the order of $\epsilon^{3}$, but we keep it in order to compare the geodesic equations, as it will be clear in section 6.

\section{\label{sec:sec3}Parametrized mapped trajectories}
Let us fix the origin of the coordinates and consider it as the center of mass (CM) of the matter 
distribution of the n-body system. Let us, also, choice that its spatial coordinates belong to the congruence of curves $C_{\mathbf {\tilde u}}$.

\begin{figure}[htp]
\begin{centering}
\includegraphics[width= 0.5\columnwidth]{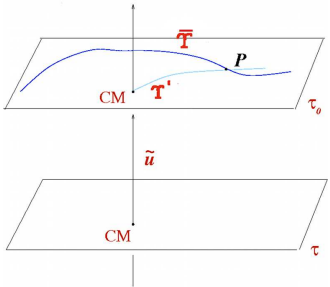}  
\caption{Mapped photon trajectory in a slice $\tau_0$ and normal 
neighbourhood of the origin of the spatial coordinates $\xi^{i}$ (CM)}
\label{f6}
\end{centering}
\end{figure}

Assume that the mapped spatial photon trajectory 
$\bar{\Upsilon}_{\bar {\ell} }$, in the slice $\tau=\tau_0$, belongs to a
{\it normal} neighbourhood of CM.
By definition then, there exist a unique 
geodesic (in any case spacelike geodesic) 
$\Upsilon '$ connecting CM to any point $\cal P$ of $\bar{\Upsilon}_{\bar {\ell} }$.
Denoting with $\sigma(\tau,{\xi}^i)$ the parameter along  
$\bar{\Upsilon}_{\bar{\ell}}$, with $\Lambda$ the parameter along each of the 
curves $\Upsilon^{'}$ stemming from CM and such that $\Lambda_{CM}=0$,
and with $\varphi$ the homeomorphism which defines the chart containing $\cal P$ in the normal 
neighbourhood of CM, we have \citep{1990recm.book.....D}: 
\[
{\cal P}= \bar{\Upsilon} (\sigma)= \Upsilon '(\Lambda) = \varphi^{-1}
({\xi}^i(\sigma)). 
\]
The coordinate $\xi^i$ can be expressed as: 
\[
 {\xi}^i =\varphi^{i}({\cal P})= 
[ \varphi \circ \bar \Upsilon (\sigma)]^i
=[ \varphi \circ \Upsilon '(\Lambda)]^i.
\]
Applying to both members of the above equation the map $\Upsilon^{' -1} \circ \varphi^{-1}$, we obtain :
     
\[
\Lambda =(\Upsilon^{' -1}\circ \varphi^{-1}) \circ 
(\varphi \circ \bar \Upsilon (\sigma)) = \Lambda(\xi^i (\sigma)),
\]
namely the $\Lambda$ implicit dependence on $\sigma$ at the point $\cal{P}$
with coordinates $\xi^i$.
Denoting as $\zeta^{i}\equiv \displaystyle{\frac{d \xi^i}{d \Lambda}}$,
the unit tangent to the curve $\Upsilon^{'}$, we define the 
length of $\Upsilon^{'}$ between CM and $\cal{P}$ as the quantity:
\begin{equation}
L = \int^{\Lambda(\xi^i (\sigma)}_0 (\zeta^{i} \, 
\zeta_{i})^{1/2} d \Lambda ' \equiv
\Lambda(\sigma).
\label{eq:length}
\end{equation}
The point of closest approach of the photon to the center of mass, will correspond to the minimum of $L$. 
Thus, differentiating (\ref{eq:length}) with respect to $\sigma$, we have
\[
\left(\frac{dL}{d\sigma}\right)_{\cal{P}}  = (\zeta_i)_{\cal{P}} \, \bar{\ell}^i,
\]
and at the point of closest approach with parameter $\hat{\sigma}$ 
\begin{equation}
(\zeta_i \, \bar{\ell}^i)_{\hat{\sigma}}=0.\
\label{eq:imp}
\end{equation}
Let us set  $\Lambda(\hat{\sigma})\equiv\hat{\xi}$ as  the impact parameter which
represents the distance from $\cal P(\hat \sigma)$ to CM; this is
a constant which labels each mapped photon trajectory on any given 
slice $\tau=\tau_0$.  
Recalling that $ \xi^i (\Lambda(\sigma))=\xi^i (\sigma) $
at any point $\cal{P}$ of intersection of the 
curves $\bar \Upsilon$ with the curve $\Upsilon^{'}$, then, at CM, where $ \xi^{i}_{CM}=  0 $ and $  \Lambda_{CM}= 0 $, 
we have to the first order in $\Lambda$:
\begin{eqnarray}
\xi^i (\Lambda + \delta \Lambda) &=& \xi^i (\Lambda_{CM})+ 
\left(\frac{d \xi^i}{d \Lambda} \right)_{\Lambda_{CM} }
\delta \Lambda+...\nonumber \\
&=& \left(\zeta^{i}_{0} \right)_{(CM \rightarrow \cal P)} \, \delta \Lambda +...
\label{eq:1exp}
\end{eqnarray}
where $\left(\zeta^{i}_{0} \right)_{(CM \rightarrow \cal P)}$ refers to the tangent 
vector at the origin of the ``unique'' spacelike geodesic connecting CM, the origin, to the point $\cal P$. From (\ref{eq:1exp}) we clearly have:
\begin{equation}
\xi^i (\Lambda)= \left(\zeta^{i}_{0}\right)_{(CM \rightarrow \cal P)} 
\int^{\Lambda}_{0} d \Lambda = \left(\zeta^{i}_{0}\right)_{(CM \rightarrow \cal P)}
\Lambda
\label{eq:sas}
\end{equation}
%
Let us again introduce a new parameter on $\bar \Upsilon$, namely
\[
\hat \tau \equiv \sigma - \hat{\sigma}
\]
then, at a point $\cal P$ on $\bar \Upsilon$ with parameter $ \hat{\sigma} + \delta \hat \tau$
\begin{equation}
\xi^i (\hat \sigma +\delta \hat \tau) = \xi^i(\hat{\sigma})  
+ \left(\frac{d \xi^i}{d \hat \tau} \right)_{\hat \tau=0} \delta \hat \tau +...
\label{eq:2exp}
\end{equation}
But from (\ref{eq:sas}), at $ \Lambda (\hat \sigma)$, it is also 
\begin{equation}
\xi^i \left(\Lambda (\hat{\sigma}) \right) = 
\left(\zeta^{i}_{0}\right)_{(CM \rightarrow P)} \hat{\xi} \equiv \hat{\xi}^i 
\label{eq:sas2}
\end{equation}  
hence, at any point $\cal P$ on $\Upsilon$, we obtain
\begin{equation}
\xi^i(\sigma)=  \hat{\xi^i} + \int_{0}^{\hat \tau} \bar{\ell^{i}} d\hat \tau^{'},
\label{eq:path}
\end{equation}  
i.e., the coordinate path of the photon on the slice $\tau_0$ as function of the two parameter $\hat{\xi}^{i}$ and $\hat \tau$.
Formula (\ref{eq:path}) has been obtained for a mapped trajectory and it contains the local spatial direction in integral form. In this sense generalizes  the parameterization of  \citet{2002PhRvD..65f4025K} (and any references therein), where, instead, the Euclidean scalar product is used (i.e. $\bar{\ell}^{i} \bar{\ell}_{i}=\delta_{ij} \bar{\ell}^{i} \bar{\ell}^{j}$). Hence, if the euclidean scalar product is applied and in case of constant light direction, equation (\ref{eq:path}) is identically the same.
However, by calculating the square modulus of (\ref{eq:path}) we get  
\begin{equation}
r^2= \xi^i  \xi_i = \hat{\xi}^2 + \hat \tau^2.
\label{eq:r-mod}
\end{equation}
The reader should bear in mind that  equation~(\ref{eq:path}) has its validity only up to $\epsilon^{2}$ and can be traced onto the hypersurface where the photon's trajectory is mapped. 

\section{\label{sec:sec4}Null geodesics in comparison}
The quantity $\bar{\ell}^{\alpha}$ (eq.~\ref{eq:ellbar}) in RAMOD is the unitary four-vector representing the \emph{local line-of-sight} of the photon as measured by the local observer $\mathbf{u}$, at rest with respect to the barycenter of the n-body system. In short, it represents a physical entity. Scope of this section is to demonstrate, first, that substituting its coordinate counterpart, within the appropriate approximations, to equation (\ref{eq:diffeqk}) of RAMOD4 is equivalent to recover light geodesic equation in the first post-Minkowskian regime adopted in \citet{2002PhRvD..65f4025K} or \citet{1999PhRvD..60l4002K}. Second, the above authors parametrize the coordinate expression of the null geodesic to solve  the light propagation problem in many physical context, as the literature largely shows, but, as far as RAMOD is concerned, only equation (\ref{eq:geodint}) of RAMOD3 recovers the parametrized one (37) in \citet{2002PhRvD..65f4025K} (equivalent to equation (19) in \citet{1999PhRvD..60l4002K}). 
Nevertheless, let us use the definition of $\mathbf{\bar{\ell}}^{\alpha}$; since it is
\[
\bar{\ell}^{i}=-\frac{k^{i}}{u_{\alpha}k^{\alpha}}\approx-\frac{k^{i}}{u^{0}k^{0}\left[-1+h_{00}+h_{0i}(k^{i}/k^{0})\right]},\]
with $u^{0}=\left(-g_{00}\right)^{-1/2}$ and $k^{i}/k^{0}=\mathrm{d}x^{i}/\mathrm{d}x^{0}$, the spatial coordinate components of $\bar{\ell}^{i}$ result (in what follows we assume $c=1$):
 \begin{eqnarray}
\bar{\ell}^{i} & = & \dot{x}^{i}\left(-g_{00}\right)^{1/2}\left(1-h_{00}-h_{0i}\dot{x}^{i}\right)^{-1}\nonumber \\
& = & \dot{x}^{i}\left(1+\frac{1}{2}h_{00}+h_{0i}\dot{x}^{i}\right)+\mathcal{O}\left(h^{2}\right).\label{eq:expr-li}
\end{eqnarray}
Let us transform equation (\ref{eq:diffeqk}) of RAMOD4 into its coordinate expression and consider only its spatial part. 
Neglecting all the contributions non linear in $h$, from (\ref{eq:expr-li}) and (\ref{eq:epsi}) the left-hand side results: 
\begin{eqnarray}
\frac{d \bar{\ell}^k}{d\sigma}&\approx& \frac{d }{d \tau} \left[ \dot{x}^k(1+\frac{1}{2} h_{00}+ h_{0i} \dot{x}^i )\right]\nonumber\\
&=& \ddot{x}^k (1+\frac{1}{2} h_{00}+ h_{0i} \dot{x}^i ) +  \nonumber \\ 
&&\dot{x}^k \left ( \frac{1}{2} \frac{d h_{00}}{d\tau } + \frac{d h_{0i}} {d\tau} \dot{x}^i + h_{0i} \ddot{x}^i \right)\nonumber\\
&=&  \ddot{x}^k + \dot{x}^k \left ( \frac{1}{2}  h_{00,i} \dot{x}^i +  \frac{1}{2}  h_{00,0} +\right.  \nonumber \\
&& \left.    h_{0i,j} \dot{x}^i \dot{x}^j +  h_{0i,0} \dot{x}^i  \right) + \mathcal{O}(h^2), \label{eq:dls/ds}
\end{eqnarray}
while the right-hand-side transforms as:
\begin{eqnarray}
&&\frac{1}{2} \dot x^k \dot x^i \dot x^j h_{ij,0} - \dot x^i \dot x^j\left(h_{kj,i}- \frac{1}{2} h_{ij,k}\right) -\nonumber \\
&&  \frac{1}{2}\dot x^k \dot x^i h_{00,i}-\dot x^i\left( h_{k0,i}+ h_{ki,0}- h_{0i,k}\right)+ \nonumber \\
&&\frac{1}{2}h_{00,k} +\dot x^{k} \dot x^{i} h_{0i, 0} - h_{k0,0}
\label{eq:dlr/ds} 
 \end{eqnarray}
Equating expressions (\ref{eq:dls/ds}) and (\ref{eq:dlr/ds}) we get the approximated null geodesic equation (40) in \citet{1999PhRvD..60l4002K} (and references therein), namely in coordinate form:
\begin{eqnarray}
\ddot{x}^k& \approx &  \frac{1}{2} h_{00,k} - h_{0k,0} - \frac{1}{2} h_{00,0} \dot{x}^k -   h_{ki,0} \dot{x}^i - \nonumber \\
&& \left(h_{0k,i} -h_{0i,k} \right) \dot{x}^i- h_{00,i} \dot{x}^k  \dot{x}^i -\nonumber\\
&&  \left(  h_{ki,j} -\frac{1}{2}h_{ij,k} \right) \dot{x}^i  \dot{x}^j + \nonumber \\
&& \left( \frac{1}{2} h_{ij,0}- h_{0i,j} \right)  \dot{x}^i \dot{x}^ j \dot{x}^ k  
\label{eq:eq-kop}.   
\end{eqnarray} 
We remind that the term $\ddot{x}^k$ is of the order of $h$ because of the order of the geodesic equation itself.  Actually, this last result should be expected, since both equations, (\ref{eq:diffeqk}) of RAMOD4 and (40) in \citet{1999PhRvD..60l4002K}, are deduced from the null geodesic (\ref{eq:null-geodesic}) in a weakly field regime. 
Once obtained this equivalence at the $h$ order by using the coordinate expression (\ref{eq:expr-li}) of $\bar \ell$, in principle one could adopt the same parameterization done, for instance, by \citet{2002PhRvD..65f4025K} (equation (36), or \cite{2007PhRvD..75f2002K}, equation (7) and reference therein) in order to find a solution of  the master equation in the RAMOD framework applicable to all of the physical cases already presented in the literature in this context. These parameters are: (i) $\tau$, defined as the Euclidean scalar product between the light ray vector and its unperturbed trajectory (eq. (13) in \cite{2002PhRvD..65f4025K}, for example), treated as a non-affine parameter, and  (ii) $ \hat \xi = \hat \xi^i  \hat \xi_i$, the constant impact parameter of the unperturbed trajectory of the light ray. 
But as section \ref{sec:sec3} shows, this kind of parameterization in RAMOD is possible only in a static space-time where null geodesic can be entirely mapped into a hypersurface simultaneous to the time of observation, i.e. at the $\epsilon^2$ level of accuracy in the linear regime.    
Then, consistently to the previous reasoning, we need to check if RAMOD3 master equation can be transformed into eq. (36) of \citet{2002PhRvD..65f4025K}. Therefore, let us adopt  the parameterization (\ref{eq:path}) in case of a constant light direction, say $\ell_{0}^{i}$; moreover, let us introduce the impact parameter $\hat \xi^{k}$ as done in \cite{1997JMP....38.2587K}, i.e. assume the equivalence of the two parameterizations as a consequence of the following change of coordinates:
\begin{eqnarray}
\hat {\xi}^k&= & P^{k}_{i} \xi^i - \ell_{0}^k \hat \tau + \Xi^{k}, \nonumber\\
\hat {\xi}^{0}&=&\hat \tau. \label{eq:coord-change}
\end{eqnarray}
In this transformation the impact parameter $\hat \xi^{k}$ is associated to the coordinate projected orthogonally (with respect to the light direction) by the operator $P^{\alpha}_{\beta}= \delta^{\alpha}_{\beta} - \ell^{\alpha} \ell_{\beta}$;  $\Xi^{k}$ represents a correction to the coordinates of the same order of the perturbation term $|h|$ of the metric, and  $d \hat \tau= d\sigma$.
Once adopted such a new coordinate system, the partial derivatives should also change according to the following rule:
\begin{eqnarray}
\frac{\partial}{\partial \xi^i}&=& \frac{\partial \hat \xi^j}{\partial \xi^i} \frac{\partial}{\partial \hat \xi^j}+  \frac{\partial \hat \xi^{0}}{\partial \xi^i} \frac{\partial}{\partial \hat \xi^{0}}=P^{j}_{i} \frac{\partial}{\partial \hat \xi^j} + \Xi^{i}_{,j}.  \label{eq:dxi}
\end{eqnarray}
The next step is to compute the new components of the metric $h_{\alpha \beta}$ according to transformations (\ref{eq:coord-change}): 
\begin{eqnarray}
h_{00}&= &\frac{\partial {\hat \xi}^0}{\partial \xi^0} \frac{\partial {\hat \xi}^0}{\partial \xi^0} \hat h_{00} + 2  \frac{\partial {\hat  \xi}^p}{\partial \xi^0} \frac{\partial {\hat \xi}^0}{\partial \xi^0} \hat h_{p0}+  \frac{\partial {\hat \xi}^p}{\partial \xi^0} \frac{\partial {\hat \xi}^q}{\partial \xi^0} \hat h_{pq} \nonumber\\
    &=& \hat h_{00} - 2 \ell_{0}^{p} \hat h_{0p}  +\ell_{0}^{p} \ell_{0}^{q}\hat h_{qp} + \mathcal{O}(h^2)\ \label{eq:g00}, 
\end{eqnarray}
\begin{eqnarray}
h_{0i}&= &\frac{\partial {\hat \xi}^0}{\partial \xi^0} \frac{\partial {\hat \xi}^0}{\partial \xi^i} \hat h_{00} +  \frac{\partial {\hat \xi}^p}{\partial \xi^0} \frac{\partial {\hat \xi}^0}{\partial \xi^i} \hat h_{p0}+\frac{\partial {\hat \xi}^0}{\partial \xi^0} \frac{\partial {\hat \xi}^q}{\partial \xi^i} \hat h_{0q} +  \nonumber \\
&&\frac{\partial {\hat \xi}^p}{\partial \xi^0} \frac{\partial {\hat \xi}^q}{\partial \xi^i} \hat h_{pq} = P^{q}_{i} \hat h_{ 0q} - \ell^{p}_{0} P^{q}_{i} \hat h_{pq} + \mathcal{O}(h^2)\ \label{eq:g0i}, 
\end{eqnarray}
and
\begin{eqnarray}
h_{ij}&= &\frac{\partial {\hat \xi}^0}{\partial \xi^i} \frac{\partial {\hat \xi}^0}{\partial \xi^j} \hat h_{00} +  \frac{\partial {\hat \xi}^p}{\partial \xi^i} \frac{\partial {\hat \xi}^0}{\partial \xi^j} \hat h_{p0}+\frac{\partial {\hat \xi}^0}{\partial \xi^i} \frac{\partial {\hat \xi}^q}{\partial \xi^j} \hat h_{0q} +  \nonumber\\
&&  \frac{\partial {\hat \xi}^p}{\partial \xi^i} \frac{\partial {\hat \xi}^q}{\partial \xi^j} \hat h_{pq}= P^{p}_{i} P^{q}_{j} \hat h_{pq} + \mathcal{O}(h^2)\label{eq:gij}.
\end{eqnarray}
We can recast the RAMOD3 master equation by applyng formula (\ref{eq:expr-li}) and keeping in mind that, all along the mapped photon trajectory, we are not allowed to neglect the derivative of the metric with respect to $\sigma$, or the new parameter $\hat \tau$, since they vary,  for each Lie-transported spatial coordinate, according the one-parameter local diffeomorphism~(\ref{eq:curveim})
\begin{eqnarray}
&&\ddot{\xi}^k  + \dot {\xi}^{k} \left(\frac{1}{2} \frac{d h_{00}}{d \sigma} \right) \approx  \nonumber \\
&&  \frac{1}{2} h_{00,k} -    \frac{1}{2}h_{00,i} \dot{\xi}^k  \dot{\xi}^i - \left(  h_{ki,j} -\frac{1}{2}h_{ij,k} \right) \dot{\xi}^i  \dot{\xi}^j, 
\label{eq:eq-ramod3-coord}
\end{eqnarray}
which is equivalent, setting $\ell_{0}^{k}$ as a constant, to:
\begin{eqnarray}
\frac {d^{2} \Xi^k}{d \hat \tau^{2}} &\approx& \frac{1}{2} \check P^q_k  h_{00,\hat q}  - \frac{1}{2} \frac{d h_{00}}{d \hat \tau}  \ell_{0}^k   -\frac{1}{2} P^{q}_{i }h_{00,\hat q} \ell_{0}^k  \ell_{0}^i - \nonumber \\
&& \left( P^{q}_{j} h_{ki,\hat q} -\frac{1}{2} P^{q}_{k} h_{ij,\hat q} \right) \ell_{0}^i  \ell_{0}^j   \label{eq:eq-ram3-der}
\end{eqnarray}
Now, considering also that $ d h_{00}/d \hat \tau= P^{q}_{i} h_{00,\hat q} \ell_{0}^{i} + h_{00, \hat \tau}$, we can eliminate the projected term along the corresponding coordinate direction and, after substituting the new metric terms, find:
 
\begin{eqnarray}
\frac {d^{2} \Xi^k}{d \hat \tau^{2}} &\approx& \frac{1}{2} \left (\hat h_{00} - 2 \ell_{0}^p \hat h_{p0} + \ell_{0}^p \ell_{0}^q \hat h_{pq} \right)_{,\hat {k}} - \nonumber \\
 && \frac{1}{2} \left (\hat h_{00}  - 2 \ell_{0}^p \hat h_{p0} + \ell_{0}^p \ell_{0}^q \hat h_{pq} \right)_{,\hat \tau}  \ell_{0}^k  \nonumber \\
 &&  - \left( \frac{1}{2}  P^p_i P^q_j \hat h_{pq} \right)_{,\hat k} \ell_{0}^i  \ell_{0}^j,  \nonumber 
\end{eqnarray}
i.e.
\begin{eqnarray}
\frac {d^{2} \Xi^k}{d \hat \tau^{2}}  &\approx& \frac{1}{2} \left (\hat h_{00} -2 \ell_{0}^p \hat h_{p0} + \ell_{0}^p \ell_{0}^q \hat h_{pq} \right)_{,\hat k} -  \nonumber \\
&& \frac{1}{2} \left (\hat h_{00}  -2 \ell_{0}^p \hat h_{p0} + \ell_{0}^p \ell_{0}^q \hat h_{pq} \right)_{,\hat \tau}  \ell_{0}^k . \label{eq:eq-ram3-der}
\end{eqnarray}
The last equation is not yet comparable to the one in \citet{2002PhRvD..65f4025K}, but we have to consider the fact that the non-diagonal term like $g_{0i}$ in the static case are null, so from (\ref{eq:g0i})
\begin{equation}
P^p_i \hat h_{0p}=\ell_{0}^q \hat P^p_i \hat h_{pq} 
\end{equation}
i.e.
 \begin{equation}
- \ell_{0}^p \ell_{0}^i \hat h_{0p}=  -\hat h_{0i} + \ell_{0}^q \hat h_{iq} - \ell_{0}^q \ell_{0}^p \ell_{0i} \hat h_{pq}.
\label{eq:goi=0}
  \end{equation}
Once replaced the right-hand-side of eq.~(\ref{eq:goi=0}) in eq.~(\ref{eq:eq-ram3-der}), we straightly get:
\begin{eqnarray}
\frac {d^{2} \Xi^k}{d \hat \tau^{2}} &\approx& \frac{1}{2} \left (\hat h_{00} -2 \ell_{0}^p \hat h_{p0} + \ell_{0}^p \ell_{0}^q \hat h_{pq} \right)_{,\hat k} - \label{eq:eq-ram3-par} \\
&&\left ( \frac{1}{2}\hat h_{00} \ell_{0}^{k} -   \hat h_{0k}  + \ell_{0}^q \hat h_{kq} -  \frac{1}{2}\ell_{0}^{k} \ell_{0}^p \ell_{0}^q \hat h_{pq} \right)_{,\hat \tau}   \nonumber
\end{eqnarray}
which is exactly the same equation (37) \citet{2002PhRvD..65f4025K}. Actually, at this stage, these authors introduce the four-dimensional isotropic vector $k^{\alpha}= (- 1, k^{i}) $ in order to get the following equation:
\begin{eqnarray}
&& \frac {d^{2} \Xi^k}{d \hat \tau^{2}} \approx  \label{eq:kop-par} \\
&& \frac{1}{2} k^{\alpha} k^{\beta} h_{\alpha \beta , \hat k} - \left ( \frac{1}{2}\hat h_{00} k^{k} + k^{\alpha} \hat h_{k \alpha} -  \frac{1}{2}k^{k} k^p 
k^q \hat h_{pq} \right)_{,\hat \tau}  \nonumber .
\end{eqnarray}
In conclusion, as far as RAMOD framework is concerned, equation~(\ref{eq:eq-ram3-par}) can be considered as an ordinary, second order differential equation for the perturbation $\Xi^k$  in variables $\hat \xi^{\alpha}$, valid in the domain where $\epsilon^{2}$ accuracy holds, i.e. where RAMOD3-like master equations must be used.  

\section{\label{sec:sec5}Body's velocity terms in the metric}
\begin{figure} 
\begin{centering}
\includegraphics[width= 0.8\columnwidth]{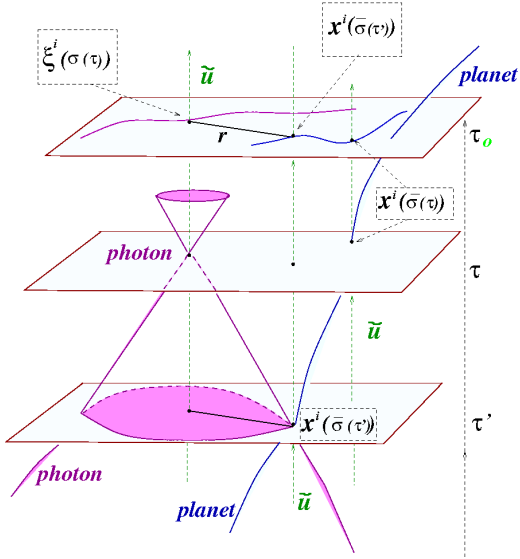}
\caption{Mapped curves related to the planet's retarded positions and retarded distance $r$. The pink line represents the mapped trajectory of the photon, here represented at time $\tau$ with a light cone. Also the coordinates of the planet $x^i(\bar {\sigma}(\tau))$ can be Lie-transported along the congruence of curves and mapped on the same slice (blue line). The contribution to the metric coefficient at each point of the photon's spatial position at coordinate time $\tau$ is determined by the source of gravity when it was located at a point on its spatial path, along the light cone, at retarded time $\tau'$; $r$, in this picture, represents the spatial coordinate distance on slice $S(\tau_0)$ between the mapped photon's position at coordinates time $\tau$ and the mapped source's location at coordinate time $\tau'$.} 
\label{fig:fig7} 
\end{centering}       
\end{figure}

The integration of the master equations requires to calculate the metric coefficients $h_{\alpha\beta}$, which depends on to the retarded distance $r_{(a)}$ as discussed in \cite{2004ApJ...607..580D} and \cite{2006ApJ...653.1552D}.
This means that we have to compute the spatial coordinate distance $r_{(a)}$ from the points on the photon trajectory to the barycenter of the a-th gravity source at the appropriate retarded time and up to the required accuracy. 
Exploiting the mapping procedure we have used to identify the spatial photon trajectory on the slice at the observation time, we can map on the same slice the trajectory of each body of the system.  We shall term $\xi^i(\sigma)$ the spatial coordinates along the spatial photon path $\hat{\mit\Upsilon}$ and $x^i(\bar\sigma)$ those on the spatial path of the a-th source, $\bar\sigma$ being the parameter along its spacetime trajectory (see figure \ref{fig:fig7}). Evidently, the metric coefficients at each point of the photon's spatial path with parameter $\sigma(\tau)$ corresponding to the photon's spatial position at the coordinate time $\tau$, are determined by each source of gravity when it was located at a point on its spatial path with parameter $\bar\sigma(\tau')$, where $\tau'$ is the retarded time: $\tau'=\tau-r$ ($c=1$). The coordinate distances is given by (we drop the suffix $(a)$):
\begin{equation}
\label{eq:erre} 
r=|\xi^i(\sigma(\tau))-x^i(\bar\sigma(\tau'))| \quad (i=1,\,2,\,3). 
\end{equation} 
Since all the functions here are smooth and differentiable, we can expand 
the coordinates of the gravitating bodies in Taylor series around their 
position at time $\tau$ as:  
\begin{equation} 
\label{eq:coorplanet} 
x^i(\bar\sigma(\tau+\delta\tau))=x^i(\bar\sigma(\tau))+\left(\frac{dx^i} 
{d\bar\sigma}\frac{d\bar\sigma}{d\tau}\right)_\tau \delta\tau+\cdots 
\end{equation} 
As reported in papers \cite{2004ApJ...607..580D, 2006ApJ...653.1552D}, the right-hand-side of (\ref{eq:erre}) is equivalent then to ($c=1$):  
\begin{eqnarray} 
\xi^i(\sigma(\tau))-x^i(\bar\sigma(\tau'))&=&\xi^i(\sigma(\tau))-
x^i(\bar\sigma (\tau))+ \label{eq:erre2}  \\
&& \int^\tau_{\tau'}
( u_\beta\bar u^\beta)^2{\rm e}^{-\psi}
\bar v^i d\tau+\cdots, \nonumber 
\end{eqnarray}
where $\bar v^\alpha$ is the spatial velocity of the source in the rest frame of the barycenter.
If we want our model be accurate to $\epsilon^3$, then it suffices  that the retarded distance $r$ contributed to the gravitational potentials, 
which we recall are at the lowest order $\epsilon^2$, at least to the order of $\epsilon$. Hence equation (\ref{eq:erre}) leads to:
\begin{equation}
\label{eq:rfinal}
r=r_0+\delta_{ij}\Delta\xi^i\int^\tau_{\tau'} \bar v^jd\zeta+{\rm O}[\epsilon^2]
\end{equation}
where
\begin{equation}
r_0=\sqrt{\delta_{ij}\Delta\xi^i\Delta\xi^j}
\label{eq:erre0}
\end{equation}
and
\begin{equation} 
\label{eq:deltax} 
\Delta\xi^i=\xi^i(\sigma(\tau))-x^i(\bar\sigma(\tau)). 
\end{equation} 
Evidently, to the order of $\epsilon^2$ (static geometry)  the contribution by  the relative velocity of the gravitating sources can be neglected. Indeed, in the static case one can further expand the retarded distance in order to keep the terms depending on the planet's velocity up to the desired accuracy. Obviously the gravitational field does not vary as photon moves from the star to the observer, since terms $g_{0i}$ are null and the time derivatives of the metric are of the order of $\epsilon^3$, therefore, the effects due body's velocity cannot be related to a dynamical change of the space-time. 
Actually, the positions of the bodies at different values of the parameter $\bar{\sigma}(\tau)$  can be recorded as subsequent snapshots into the mapped trajectories and deduced as postponed corrections in the reconstruction of the photon's path.   
\section{\label{sec:comments}Conclusions}
Most of the information,  from the Universe around us, comes in form of electromagnetic waves. While propagating, light interacts with the curvature generated by the encountered mass. Then, it is of utmost importance to reconstruct the whole light path precisely as much as the varying curvature. For such a reason, the covariant RAMOD framework aims to use and preserve as much as possible all the physical quantities needed to reconstruct the history of the photon.
As a matter of fact, modelling light propagation is intrinsically connected to the identification of the geometry where photons naturally move. This process can substantially influence the solution of such a problem, since, according to the physical environment where observations take place, appropriate metric terms should be retained or ruled out in the affine connection that contributes to make a null geodesic explicit and integrable.

The result reported here, in the case of RAMOD-like models, represents an improvement in the understanding of how to handle the null geodesic within the accuracy requirements due to the geometry, i.e. the physics of a n-body bound system. 
To avoid confusion, the reader should bear in mind that the present work does not deal with the description of the n-body dynamics; it considers, instead, how the set-up of the space-time for light propagation and detection changes without or with the vorticity term, being the latter related to the congruence of curves that foliates the space-time and allows to define a Lie-transported coordinate system.  The vorticity term cannot be neglected at the the order of $\epsilon^{3}$: the possibility to ignore it locally  can be applied only to a small neighborhood with respect to the scale of the vorticity itself. In other words, we have to scrutinize if a measurement can be considered local or not with respect to the curvature \citep{2010ToM.book.....D}.

Within the scale of the Solar System, then, there is none slice which extends from the observer up to the star that emits light. This justifies why only RAMOD3 recovers the parametrization used in \citet{1999PhRvD..60l4002K}, namley when the geometry allows to define a rest-space of an observer everywhere. 
In this regard, this paper proves that in RAMOD3, i.e., a static space-time model,  the parameterization (\ref{eq:path}) is appropriate to solve the light ray tracing problem. Formula (\ref{eq:path})  has its validity in $\epsilon^2$ regime for a mapped trajectory and it contains the local spatial direction in integral form. It generalizes the parameterization used in \citet{1999PhRvD..60l4002K} and, consistently, the RAMOD3 master equations, once converted into a coordinate form, recover the analytical linearized case discussed by Kopeikin and others. Consequentely, those master equations are solvable with the same technique and may include any detectable relativistic effects associated to the light propagation in that context, but only at  $\epsilon^2$ level of accuracy. 

Instead, the RAMOD4 model, i.e. the case of a dynamical space-time, fully preserves the "active" contents of gravity. Let us remind that formulas (\ref{eq:diffeq0}) and (\ref{eq:diffeqk}) are not contemplated in other approaches: only equation (\ref{eq:diffeqk}), referred to the spatial components, can be reduced, once transformed into a coordinate form, to the equation of propagation of photons in a weak gravitational field in the first pM approximation, as demonstrated in this paper. Actually, RAMOD4 master equations define a set of nonlinear differential equation for the four-vector $ \mathbf{ \bar \ell}$, the {\it spatial} local-line-of-sight as measured in the rest-space of a local barycentric observer which feels the gravitational field according to (\ref{eq:u}). This family of observers can defined in each event of the space-time, then also at the observer's location. 
Methods to solve light tracing analytically in RAMOD4  are under investigations; to this class of methods belongs the semi-anlytical solution of \citet{2006CQGra..23.5467D}. 

A new  analytical solution for the RAMOD master equations will surely represent a substantial improvement for the light propagation problem, since RAMOD ''recipe'', based on General Relativity,  can tackle light tracing and detection to any order of accuracy and, even more important, consistently with the space-time geometrical set-up. This fact reflects the main purpose of the RAMOD approach which is to express the null geodesic, or the equivalent master equations, by all the physical quantities entering the process of observation, in order  to entangle globally all the possible interactions of light with the background geometry. Keeping the physical aspects of the problem guarantees the consistency of the measured physical effects with the intrinsic accuracy of the space-time. A preferred role is played in RAMOD by the local-line-of-sight,  the main unknown of the master equations, which represents what {\it locally} the observer measures of the collected light. At the time of observation it represents the boundary condition to solve uniquely the light path. In a second step, once the light is traced back, any hidden subtle relativistic effects can be split by the global solution case by case, also according to the different level of accuracy needed, and finally quantified by a suitable coordinates system.

As a concluding remark, by recovering results already presented in the literature, what this paper clearly establishes is the full potential of the RAMOD construct: its general covariant formulation represents a well defined framework where any desired advancement in the light tracing problem and its subsequent detection as physical measurement can be contemplated.

\begin{acknowledgments}
This work is supported by the ASI contracts I/037/08/0 and I/058/10/0.
\end{acknowledgments}
\appendix
\section{Appendix}
Let us choose the coordinate system (\ref{eq:coor1})  adapted to the spatial 
hypersurfaces $\tau=constant$ and designate two hypersurfaces $\tau_1$ and  $\tau_2$.
In order to compare proper length and proper time on them, we have to evaluate the line element, that is:
\begin{eqnarray}
ds^2 &= &{g}_{\alpha \beta} d x^{\alpha} d x^{\beta}= \nonumber \\
&=&{g}_{00}d x^{0} d x^{0}+ 2 {g}_{0i} d x^{0} 
d x^{i}+{g}_{ij} d x^{i} d x^{j}.
\label{eq:line}
\end{eqnarray}
   
We know that:
\begin{equation}
\label{eq:ucheck}
\begin{cases}
{u}_{\alpha}(x^0,x^i)= -\delta^{0}_{\alpha} {\rm e}^{\psi} \cr 
{u}^{\alpha}(x^0,x^i)= -{g}^{0 \alpha}{\rm e}^{\psi} & {}
\end{cases}
\end{equation}
so 

\[
{u}_i={g}_{i\beta}{u}^{\beta}={g}_{i0} 
{u}^{0}+{g}_{ik}{u}^{k}=0
\]
\[
\Longrightarrow {g}_{i0}= -\frac{{g}_{ik} {u}^{k}}
{{u}^{0}}
\]
and
\[
{u}^{0}= -{\rm e}^{\psi} {g}^{00}.
\]
Substituting the latter expressions in (\ref{eq:line}), and 
adding and subtracting the same quantity 
$\displaystyle {\frac{{u}^{i} {u}^{j}}{({u}^{0})^2} 
(d x^0)^2}$, we get:

\begin{eqnarray}
ds^2 &=& (dx^0)^2 \Big[{g}_{00} - {g}_{ij}
\frac{{u}^{i}{u}^{j}}{ ({u}^{0})^2} \Big]+ \nonumber \\
&&  {g}_{ij} \Big[\Big(d x^{i} - \frac{{u}^{i}}{{u}^{0}} d x^0 \Big)\Big]
\Big[\Big(d x^{j}- \frac{{u}^{j}}{{u}^{0}} d x^0\Big)\Big] 
\end{eqnarray}
Manipulating the condition ${u}_{\alpha}{u}^{\alpha}=-1$, 
we find:
\[
{g}_{\alpha \beta} {u}^{\alpha} {u}^{\beta}=
-{g}_{00}{g}^{00} -g_{ij}{u}^{i}{u}^{j}=-1
\]
\[
\Longrightarrow  {g}_{ij}{u}^{i}{u}^{j}=
1 -{g}_{00}{g}^{00}. 
\]
Hence, denoting $N^i=\displaystyle {\frac{{u}^{i}}{{u}^{0}}}$
({\it shift factor}) and $N=\displaystyle {\pm \frac{1}{{u}^{0}}}$
({\it lapse factor}), equation (\ref{eq:line}) becomes:
\begin{equation}
ds^2 = -(N d x^0)^2 +  {g}_{ij}
\Big( d x^{i} - N^i d\xi^0 \Big) \Big(d x^{j}- N^j dx^0 \Big).
\label{eq:nline}
\end{equation}
%

%
\section{Detailed calculations for formula (\ref{eq:geoC})} 
\label{app:appB}

In what follows we drop the tilde. 
Let ${\bf k}$ be the null tangent vector to the photon's geodesic.
It can be expressed as (\ref{eq:ell} or \ref{eq:hatell}):
\begin{equation}
k^{\alpha}=l^{\alpha}-(u \mid k) u^{\alpha}.
\label{eq:kvsl}
\end{equation}
Then, the geodesic equation can be derivate, step by step, as it follows:
\begin{eqnarray}
\frac{d [l^{\alpha}-(u \mid k) u^{\alpha}]}{d \lambda}+
\Gamma^{\alpha}_{\beta \gamma}[l^{\beta}-(u \mid k)u^{\beta}]
[l^{\gamma}-(u \mid k)u^{\gamma}]=0   \nonumber 
\end{eqnarray}
\begin{eqnarray}
\frac{d l^{\alpha}}{d \lambda}-(u \mid k)\frac{du^{\alpha}}
{d \lambda}- u^{\alpha} \Big(\frac{D u}{D\lambda} \mid k \Big) &+& \nonumber \\ 
\Gamma^{\alpha}_{\beta \gamma}[l^{\beta}l^{\gamma}-(u \mid k)
l^{\beta}u^{\gamma}-(u \mid k) l^{\gamma}u^{\beta}+
(u \mid k)^2 u^{\beta}u^{\gamma}]=0 \nonumber
\end{eqnarray}

\begin{eqnarray}
\frac{d l^{\alpha}}{d \lambda}-(u \mid k)\frac{du^{\alpha}}
{d \lambda}- u^{\alpha} [k^{\beta}k^{\tau} \nabla_{\tau}u_{\beta}] &+&  \nonumber \\
\Gamma^{\alpha}_{\beta \gamma}[l^{\beta}l^{\gamma}-(u \mid k)
(l^{\beta}u^{\gamma}+l^{\gamma}u^{\beta})+ (u \mid k)^2 u^{\beta}u^{\gamma}]=0  \nonumber 
\end{eqnarray}
where
\[
k^{\beta}k^{\tau} \nabla_{\tau}u_{\beta}= l^{\beta}l^{\tau} 
\nabla_{\tau}u_{\beta}-(u \mid k) l^{\beta}\dot{{u}_{\beta}},
\]
and the symbol $D \mathbf u/D \lambda$ is the absolute derivative of $\mathbf u$ along the curve with parameter $\lambda$ and dots indicate derivative with respect to the parameter of the curve.  
In fact:
\[
u^{\beta}\dot{u_{\beta}}=0
\]
and
\[
u^{\beta} l^{\tau} \nabla_{\tau}u_{\beta}=0
\]
because of condition $u_{\alpha}u^{\alpha}=-1$. 
Let us trasform the affine parameter $\lambda$ of the geodesic into the parameter $\sigma$
\[
d \sigma= -(u \mid k) d \lambda.
\]  
Therefore, the derivative of \ref{eq:kvsl} results as:  
\[
\frac{d l^{\alpha}}{d \lambda} =  \frac{d [-(u \mid k) \bar{l}^{\alpha}]} 
{d \sigma} \frac{d \sigma}{d \lambda},
\]  
i.e.
\begin{eqnarray*}
\frac{d l^{\alpha}}{d \lambda}= 
(u \mid k)^2 \frac{d \bar{l}^{\alpha}}{d \sigma}+(u \mid k) \bar{l}^{\alpha}
\left( \frac{D u}{ D \sigma} \mid k \right)&=& \\
(u \mid k)^2 \frac{d \bar{l}^{\alpha}}{d \sigma}+(u \mid k) \bar{l}^{\alpha}
( \bar{k}^{\tau} \nabla_{\tau} u_{\beta} k^{\beta}) &=&\\
 (u \mid k)^2 \frac{d \bar{l}^{\alpha}}{d \sigma}-(u \mid k)^2 
\bar{l}^{\alpha} [ (\bar{l}^{\tau} + u^{\tau}) \nabla_{\tau} u_{\beta} 
\bar{k}^{\beta}]&=& \\
 (u \mid k)^2 \frac{d \bar{l}^{\alpha}}{d \sigma}-(u \mid k)^2 
\bar{l}^{\alpha} [(\bar{l}^{\tau} + u^{\tau})(\bar{l}^{\beta} + u^{\beta})
\nabla_{\tau} u_{\beta}]&=&\\
 (u \mid k)^2 \frac{d \bar{l}^{\alpha}}{d \sigma}-(u \mid k)^2 
\bar{l}^{\alpha} (\bar{l}^{\tau}\bar{l}^{\beta}\nabla_{\tau} u_{\beta}+
\bar{l}^{\beta} \dot{u}_{\beta}). 
\end{eqnarray*}
So the above geodesic equation becames:
\[
(u \mid k)^2 \frac{d \bar{l^{\alpha}}}{d \sigma}+
(u \mid k)^2 \frac{d u^{\alpha}}{d \sigma}- 
(\bar{l^{\alpha}} + u^{\alpha}) [\bar{l}^{\beta}\bar{l}^{\tau} 
\nabla_{\tau}u_{\beta}
+ \bar{l}^{\beta} \dot{u}_{\beta}] +
\]
\[ 
(u \mid k)^2 \Gamma^{\alpha}_{\beta \gamma}[\bar{l}^{\beta}\bar{l}^{\gamma}
+(\bar{l}^{\beta}u^{\gamma}+\bar{l}^{\gamma}u^{\beta})+
u^{\beta}u^{\gamma}]=0
\]
 
Finally, dividing each member by $(u \mid k)^2$ we  
find the geodesic equation (\ref{eq:geoC}).  

%

\section{The case of null expansion}
\label{app:appC}
By imposing the free expansion condition~\footnote{Round brackets means that the tensor is symmetric with respect the pair of index between the brackets and in order to simply the notations we keep the standard symbol for partial derivative}
\begin{equation}
P_{\rho}^{\alpha} P_{\sigma}^{\beta}
\bigtriangledown_{( \alpha}u_{\beta )}=\Theta_{\rho \sigma} \ell_{\rho}^{\alpha} \ell_{\sigma}^{\beta}=0,
\label{eq:expnull} 
\end{equation}
where
\[
P_{\rho}^{\alpha}= \delta_{\rho}^{\alpha}+ u_{\rho}u^{\alpha},
\]
\[
P_{\sigma}^{\beta}=\delta_{\sigma}^{\beta}+ u_{\sigma}u^{\beta},
\]
\[
{u}^{\alpha}={\rm e}^{\varphi} \delta^{\alpha}_{0},
\]
\[
{u}_{\alpha}={\rm e}^{\varphi}{g}_{0 \alpha},
\]
\[
g_{\alpha \beta}= \eta_{\alpha \beta}+h_{\alpha \beta},
\]
\[
\frac{1}{g_{00}}= -\frac{1}{1-h_{00}}\approx -(1+h_{00}),
\]
and
\[
\bigtriangledown_{\alpha} u_{\beta}= {\rm e}^{\phi}
\partial_{( \alpha}g_{\beta) 0}
-\frac{\partial_{(\alpha} g_{00}}{2 g_{00}} u_{\beta) } 
- \Gamma^{\tau}_{\alpha \beta} u_{\tau},  
\]
we find
\begin{eqnarray}
\Theta_{\rho \sigma} &=& P_{\rho}^{\alpha} P_{\sigma}^{\beta}
\bigtriangledown_{( \alpha} u_{\beta )}  \nonumber \\
&=&{\rm e}^{\varphi}\Big[(\delta_{\rho}^{\alpha} + u_{\rho}u^{\alpha})
(\delta_{\sigma}^{\beta}+  
u_{\sigma}u^{\beta}) \Big(\partial_{( \alpha}g_{\beta ) 0} 
  -\Gamma^{\tau}_{\alpha \beta} g_{0 \tau}\Big) \Big]  \nonumber 
\end{eqnarray}
because of
\[
P_{\rho}^{\alpha} P_{\sigma}^{\beta}
\frac{\partial_{(\alpha} g_{00}}{2 g_{00}} u_{\beta)}=0.
\] 
After replacing for $ \Gamma^{\tau}_{\alpha \beta}$ its expression with the metric coefficients
\[
\frac{1}{2}g^{\lambda \tau}
(\partial_{\alpha}g_{\lambda \beta}+\partial_{\beta}g_{\lambda \alpha}-
\partial_{\lambda}g_{\alpha \beta})
\]
equation \ref{eq:expnull} results term by term: 
\begin{eqnarray}
P_{\rho}^{\alpha} P_{\sigma}^{\beta}
({\rm e}^{\varphi} \partial_{(\alpha}g_{\beta ) 0})&=&
{\rm e}^{\varphi}\Big[ \partial_{(\rho}g_{\sigma) 0}+
{\rm e}^{2\varphi}g_{0 ( \rho}\partial_{|0|}g_{\sigma) 0} \\
&+& {\rm e}^{2\varphi}g_{0 (\sigma} \partial_{\rho)}g_{00}+
{\rm e}^{4\varphi}g_{0 (\sigma}g_{\rho) 0}\partial_{0}g_{00}\Big] \nonumber 
\label{eq:esp1}
\end{eqnarray}

\begin{eqnarray}
\label{eq:esp2}
P_{\rho}^{\alpha} P_{\sigma}^{\beta}\Gamma^{\tau}_{\alpha \beta}
u_{\tau}&=& -\frac{{\rm e}^{\varphi}}{2}\partial_{0}g_{\rho \sigma}
+ {\rm e}^{\varphi}\partial_{(\rho}g_{\sigma)0} \\ 
&+& {\rm e}^{3 \varphi}
g_{0 (\rho} \partial_{\sigma)} g_{00} +\frac{1}{2}{\rm e}^{5 \varphi} 
g_{0 (\sigma} g_{\rho)0} \partial_{0} g_{00}. \nonumber
\end{eqnarray}

Finally, by summing terms (\ref{eq:esp1}) and (\ref{eq:esp2}), we obtain:
\[
\Theta_{\rho \sigma}= \frac{1}{2}{\rm e}^{\varphi}\partial_{0}g_{\rho \sigma}+
{\rm e}^{3\varphi}g_{0 ( \rho}\partial_{|0|}g_{\sigma)0}+\frac{1}{2}
{\rm e}^{5\varphi}g_{0 (\sigma} g_{\rho)0} \partial_{0}g_{00}.
\]
It is easy to see that for $\rho=i$ and $\sigma=j$ the last equation becomes: 
\[
\Theta_{i j }= \frac{1}{2}{\rm e}^{\varphi}\partial_{0}g_{ij},
\] 
implying for condition \ref{eq:expnull} 
\begin{equation}
 \partial_{0}g_{ij} = 0.
\label{eq:dogij}
\end{equation}
Instead, for $\rho=0$ and $\sigma=i$:
\begin{eqnarray}
\Theta_{0 i}&=& \frac{1}{2}{\rm e}^{\varphi}\partial_{0}g_{0 i}+
\frac{1}{2}{\rm e}^{3\varphi} \left( g_{0 0} \partial_{0}g_{i 0} + g_{0 i} \partial_{0}g_{0 0}\right) \\
&+&
\frac{1}{4} {\rm e}^{5\varphi} \left(g_{0 0} g_{i0}+g_{0 i} g_{00}\right)
\partial_{0}g_{00} \nonumber \\
&=& \frac{1}{2}{\rm e}^{\varphi}\partial_{0}g_{0 i}+
\frac{1}{2}{\rm e}^{3\varphi} g_{0 0} \partial_{0}g_{0i} \nonumber \\
&=&  \frac{1}{2}{\rm e}^{\varphi} \partial_{0}g_{0 i} (1+ 
{\rm e}^{2\varphi} g_{0 0})=0 \nonumber 
\end{eqnarray}
\begin{equation}
\Longrightarrow \partial_{0}g_{0 i}=0.
\label{eq:dogoi}
\end{equation}

And, finally, for $\rho=0$ and $\sigma=0$: 
\begin{eqnarray}
\Theta_{0 0}&=&{\rm e}^{\varphi}\partial_{0}g_{0 0} \left[ \frac{1}{2}
+{\rm e}^{2\varphi}g_{0 0}+ \frac{1}{2}{\rm e}^{4\varphi} (g_{0 0})^2\right]
\nonumber \\
&=&  \frac{1}{2}{\rm e}^{\varphi}\partial_{0}g_{0 0}\left( 1
+g_{0 0}\right)^2 \nonumber
\end{eqnarray}
by which, again, from equation \ref{eq:expnull} we get: 
\begin{equation}
\Longrightarrow \partial_{0}g_{0 0}=0. 
\label{eq:dogoo}
\end{equation}


\section{Calculations for formula (\ref{eq:geodint})}
\label{app:appD}
The spatial null geodesic with respect to the expansion-free congruence 
$C_{\tilde {\mathbf u}}$ is: 
\[
\frac{d \bar{l}^{\alpha}}{d \sigma} +\frac{d \tilde{u}^{\alpha}}{d \sigma} -
 (\bar{l}^{\alpha} +\tilde u^{\alpha}) (\bar{l}^{\beta} \dot{\tilde u}_{\beta})+
\Gamma^{\alpha}_{\beta \gamma}(\bar{l}^{\beta} + \tilde u^{\beta})  
(\bar{l}^{\gamma} + \tilde u^{\gamma})=0
\]

To the first order in $h$ we have:
\begin{enumerate}
\item 
\[
\tilde u^{\alpha}= \frac{\delta^{\alpha}_{0}}{\sqrt{-g_{00}}} 
\approx \delta^{\alpha}_{0} (1+\frac{1}{2} h_{ 00})\approx
\] 
\[
\delta^{\alpha}_{0}(1+\frac{1}{2}h_{00}+ O(h^2));
\]
\item
\[
\frac{d \tilde u^{\alpha}}{d \sigma}= \bar{l}^{\beta} \partial_{\beta} \tilde u^{\alpha}=
 \bar{l}^{i} \partial_{i} \tilde u^{\alpha}\approx
\]
\[
\bar{l}^{i}\delta^{\alpha}_{0}(1+\frac{1}{2} h_{00})
(\partial_{i}h_{00})(1+h_{00})\approx \delta^{\alpha}_{0} 
\frac{1}{2}(\bar{l}^{i}\partial_{i}h_{00}) ;
\]

\item
\[
\dot{\tilde u}_{\beta}=\tilde u^{\tau} \nabla_{\tau}\tilde u_{\beta}\approx 
\delta^{\tau}_{0} \Big(1+\frac{1}{2}h_{00}\Big)\nabla_{\tau}\tilde u_{\beta}
\]

\[
\nabla_{\tau} \tilde u_{\beta}=
\frac{g_{\beta 0}}{2g_{00}\sqrt{-g_{00}} } \partial_{\tau}(- g_{00})+
\frac{1}{\sqrt{-g_{00}}}\partial_{\tau} g_{\beta 0} -
\Gamma^{\lambda}_{\tau \beta} \frac{g_{\lambda 0}}{\sqrt{-g_{00}}} \approx
\]
\[
 \Big(1+\frac{1}{2}h_{00}\Big)\Big[\frac{1}{2} (1 + h_{00})
(\eta_{\beta 0}+ h_{\beta 0})(\partial_{\tau}h_{00})+
\partial_{\tau} h_{\beta 0}- \Gamma^{\lambda}_{\tau \beta}\eta_{\lambda 0}\Big]
\approx
\]
\[
\frac{1}{2}\eta_{\beta 0}\partial_{\tau} h_{00}+
\partial_{\tau} h_{\beta 0}-  \Gamma^{\lambda}_{\tau \beta}\eta_{\lambda 0}
 + O(h^2)
\]
\[
\dot{\tilde u}_{\beta}=\tilde u_{\tau} \nabla_{\tau}\tilde u_{\beta}\approx 
\frac{1}{2}\eta_{\beta 0}\partial_{0} h_{00}+\partial_{0} h_{\beta 0}+
 \Gamma^{0}_{0 \beta}.
\]
\end{enumerate}
Adding all the above terms, the spatial null geodesic becomes, 
to the first order in $h$:
\begin{eqnarray}
\frac{d \bar{l}^{\alpha}}{d \sigma} &+& 
\frac{1}{2}\Big(\bar{l}^{i}\partial_{i}h_{00}\Big)\delta^{\alpha}_{0}-
\Big[\bar{l}^{\alpha}+\delta^{\alpha}_{0}(1+ \frac{1}{2}  h_{00})\Big] \nonumber \\
&&\Big[\bar{l}^{\beta}(\frac{1}{2}\eta_{\beta 0}\partial_{0} h_{00}+
\partial_{0} h_{\beta 0}+ \Gamma^{0}_{0 \beta})\Big]+
\end{eqnarray}
\[
\Gamma_{\beta \gamma}^{\alpha}
\Big[\bar{l}^{\beta}+\delta^{\beta}_{0}(1+ \frac{1}{2}  h_{00})\Big]
\Big[\bar{l}^{\gamma}+\delta^{\gamma}_{0}(1+ \frac{1}{2}  h_{00})\Big]=0
\]
i.e.
\[
\frac{d \bar{l}^{\alpha}}{d \sigma}+
\frac{1}{2}(\bar{l}^{i}\partial_{i}h_{00})\delta^{\alpha}_{0}-
\Big[\bar{l}^{\alpha}+\delta^{\alpha}_{0}(1+ \frac{1}{2}  h_{00})\Big]
\Big[\bar{l}^{i}(\partial_{0}h_{i0} +\Gamma_{0i}^{0})\Big]+
\]
\[
\Gamma_{ij}^{\alpha}\bar{l}^{i}\bar{l}^{j} + 
\Gamma_{i0}^{\alpha}\bar{l}^{i} (1+ \frac{1}{2}  h_{00})
+\Gamma_{0j}^{\alpha}\bar{l}^{j} (1+ \frac{1}{2}  h_{00})+
\Gamma_{00}^{\alpha}(1+ h_{00})=0
\]
where we have, at the first order in $h$:
\[
\Gamma_{\beta \gamma}^{\alpha}\approx \frac{1}{2} \eta^{\alpha \lambda}
(\partial_{\beta}h_{\lambda \gamma}+\partial_{\gamma}h_{\lambda \beta}-
\partial_{\lambda}h_{\beta \gamma})
\]
and
\[
h_{0i}=0
\]
\[
\partial_{0}h_{ij}=0.
\]
Now, for $\alpha=0$:
\[
\frac{1}{2}\bar{l}^{i}\partial_{i}h_{00}-
\Big(1+ \frac{1}{2}  h_{00}\Big) \Big(\bar{l}^{i}\partial_{0}h_{0i}
-\frac{1}{2}\bar{l}^{i}\partial_{i}h_{00} \Big)
-\frac{1}{2} \bar{l}^{i}\partial_{i} h_{00}
\]
\[
-\frac{1}{2}\bar{l}^{j}(\partial_{j}h_{00})
-\frac{1}{2}\partial_{0}h_{00}=0
\]
\[
\Longrightarrow \partial_{0}h_{00}=0,
\] 
and for $\alpha=k$ ($k=1, 2, 3$):
\[
\frac{d \bar{l}^{k}}{d \sigma}-
\bar{l}^{k}\Big[-\frac{1}{2}(\bar{l}^{i})\partial_{i}h_{00}\Big]+
\frac{1}{2}\eta^{k \lambda}\Big[(\partial_{i} h_{\lambda j}+\partial_{j} 
h_{\lambda i}- \partial_{\lambda} h_{ij})(\bar{l}^{i}\bar{l}^{j})+
\]
\[
 (\partial_{i} h_{\lambda 0}+ \partial_{0} h_{\lambda i}-
\partial_{\lambda} h_{i0}) \bar{l}^{i} +
(\partial_{0} h_{\lambda j}+ \partial_{j} h_{\lambda 0}-
\partial_{\lambda} h_{0j})\bar{l}^{j}+
\]
\[
\partial_{0} h_{\lambda 0}+ 
\partial_{0} h_{\lambda 0}-\partial_{\lambda} h_{00}\Big]=0.
\]
The terms like $\partial_{i} h_{k 0}$ are null because in our system
of coordinates $g_{\alpha 0}=0$ and $\partial_{0} h_{k i}=0$ for equation
(\ref{eq:dogij}). So, finally:
\begin{eqnarray}
\frac{d \bar{l}^{k}}{d \sigma}&+&
\bar{l}^{k}\Big(\frac{1}{2}\bar{l}^{i}\partial_{i}h_{00}\Big) \\
&+& \frac{1}{2}\eta^{k\lambda} (\partial_{i} h_{\lambda j}+ \partial_{j} 
h_{\lambda i}-\partial_{\lambda} h_{ij})(\bar{l}^{i}\bar{l}^{j})
-\frac{1}{2}\eta^{k \lambda}\partial_{\lambda} h_{00}=0. \nonumber
\end{eqnarray}
%


\bibliography{mybibl}


\end{document}